\documentclass{aa}  

\usepackage{graphicx}
\usepackage{txfonts}
\usepackage{rotating}
\usepackage{amsmath}
\usepackage{longtable}
\usepackage{tablefootnote}
\usepackage{subcaption}
\usepackage{float}
\usepackage{amssymb}
\usepackage{xcolor}
\usepackage{booktabs}
\usepackage[version=3]{mhchem}
\usepackage{mathrsfs}
\usepackage{multirow,bigdelim,dcolumn,booktabs}

\newif\ifAMStwofonts

\newcommand{\krome}{{\sc Krome}}
\newcommand{\crir}{\zeta_{\rm 2}}
\newcommand{\X}{{\tt X}}
\newcommand{\LL}{\mathscr{L}}
\newcommand{\DD}{{\rm D}}
\newcommand{\tte}{{\boldsymbol \theta}}
\newcommand{\ppc}{P_{\rm C}}
\newcommand{\ppcx}{P_{\rm C,{\tt X}}}
\newcommand{\ppm}{P_{\rm M}}
\newcommand{\ppl}{P_{\rm L}}
\newcommand{\Dtp}{\Delta t_{\rm phase}}
\newcommand{\tA}{t_{\rm A}}
\newcommand{\fA}{f_{\rm A}}

\newcommand{\tmsf}{t_{\rm MSF}}

\newcommand{\hhco}{H$_2$CO}
\newcommand{\chhhcch}{CH$_3$CCH}
\newcommand{\chhhcn}{CH$_3$CN}
\newcommand{\chhhoh}{CH$_3$OH}
\newcommand{\msun}{{\rm M}_\odot}

\begin{document} 

   \title{Establishing the evolutionary timescales of the massive star formation process through chemistry}

   \author{G.~Sabatini\inst{1,2,3}\fnmsep\thanks{Marco Polo fellowship of the University of Bologna;}
          \and
          S.~Bovino\inst{3}
          \and
          A.~Giannetti\inst{2}
          \and
          T.~Grassi\inst{4}
          \and
          J.~Brand\inst{2}
          \and
          E.~Schisano\inst{5}
          \and
          F.~Wyrowski\inst{6}
          \and\\
          S.~Leurini\inst{7}
          \and
          K.~M.~Menten\inst{6}
          }

      \institute{Dipartimento di Fisica e Astronomia ``Augusto Righi'', Universit\'a di Bologna, Via Gobetti
                 93/2, I-40129 Bologna, Italy\\ \email{giovanni.sabatini@inaf.it}
         \and
                 INAF - Istituto di Radioastronomia - Italian node of the European ALMA Regional Centre (It-ARC), Via Gobetti 101, 40129 Bologna, Italy
         \and
                 Departamento de Astronom\'ia, Facultad Ciencias F\'isicas y Matem\'aticas, Universidad de Concepci\'on, Av. Esteban Iturra s/n Barrio Universitario, Casilla 160, Concepci\'on, Chile
         \and 
                 Universit\"{a}ts-Sternwarte M\"{u}nchen, Scheinerstr. 1, D-81679 M\"{u}nchen, Germany
         \and
                 INAF – Istituto di Astrofisica e Planetologia Spaziali (IAPS), via Fosso del Cavaliere 100, 00133 Roma, Italy
         \and
                 Max-Planck-Institut f\"ur Radioastronomie, Auf dem H\"ugel, 69, 53121, Bonn, Germany  
         \and
                 INAF – Osservatorio Astronomico di Cagliari, Via della Scienza 5, 09047, Selargius (CA), Italy
             }

   \date{Received 1 February 2021 / Accepted 31 May 2021}

 
  \abstract
   {Understanding the details of the formation process of massive (i.e. M$\gtrsim$8-10M$_\odot$) stars is a long-standing problem in astrophysics. They form and evolve very quickly, and almost their entire formation process takes place deeply embedded in their parental clumps. Together with the fact that these objects are rare and at a relatively large distance, this makes observing them very challenging.}
   {We present a method for deriving accurate timescales of the evolutionary phases of the high-mass star formation process.}
   {We modelled a representative number of massive clumps of the ATLASGAL-TOP100 sample that cover all the evolutionary stages. The models describe an isothermal collapse and the subsequent warm-up phase, for which we followed their chemical evolution. The timescale of each phase was derived by comparing the results of the models with the properties of the sources of the ATLASGAL-TOP100 sample, taking into account the mass and luminosity of the clumps, and the column densities of methyl acetylene (\chhhcch), acetonitrile (\chhhcn), formaldehyde (\hhco), and methanol (\chhhoh).}
   {We find that the molecular tracers we chose are affected by the thermal evolution of the clumps, showing steep ice evaporation gradients from 10$^3$ to 10$^5$ AU during the warm-up phase. We succeed in reproducing the observed column densities of \chhhcch~and \chhhcn, but \hhco~and \chhhoh~agree less with the observed values. The total (massive) star formation time is found to be $\sim$5.2$\times 10^5$ yr, which is defined by the timescales of the individual evolutionary phases of the ATLASGAL-TOP100 sample: $\sim$5~$\times~10^4$ yr for  70-$\mu$m weak, $\sim$1.$2\times10^5$ yr for mid-IR weak, $\sim$2.$4\times10^5$ yr for mid-IR bright, and $\sim$1.$1\times10^5$ yr for HII-region phases.}
   {With an appropriate selection of molecular tracers that can act as chemical clocks, our model allows obtaining robust estimates of the duration of the individual phases of the high-mass star formation process. It also has the advantage of being capable of including additional tracers aimed at increasing the accuracy of the estimated timescales.}


   \maketitle
%

\section{Introduction}\label{sec1:intro}

Although high-mass stars (i.e. M$\gtrsim$8-10M$_\odot$) are much rarer than their less massive counterparts, they have a critical effect on the physico-chemical characteristics of the interstellar medium (ISM). They also play an important role in the evolution of the host galaxies \citep[e.g.][]{Kennicutt05}. Massive stars are responsible for the production of large amounts of the $\alpha$-elements involved in the formation of complex molecules \citep{Woosley95, Kobayashi20} that are created during the final stages of their life-cycle through core-collapse supernovae (e.g. \citealt{Smartt09}). They also dominate the energy budget in their immediate surroundings by stirring, heating, and ionising the gas, and they affect the chemical evolution of the ISM as well as the star and planet formation process (e.g. \citealt{Elmegreen98, Bally05, Adams10}). However, observing massive stars is challenging as they evolve quickly, and the initial stages of their formation process take place when their progenitors are still embedded in the parental clump \citep[][for reviews on this topic]{Zinnecker&Yorke07, Motte18}. In spite of the efforts made in recent years to understand the formation of massive stars, some fundamental questions still remain unanswered. One of these is related to the timescales of the various (evolutionary) phases of their formation process, which would provide crucial information to distinguish among competing star formation theories \citep[e.g.][]{McKee02, Mouschovias06, Hartmann12, Tige17, Padoan20} and to reach a comprehensive view of the chemical evolution of the gas that resides in the parental clumps of massive stars.\\ 
\indent Timescales are usually derived by applying statistical approaches (e.g. \citealt{Wood89, Davies11, Mottram11, Battersby17, Tige17}; see also \citealt{Motte18}). Typical values for the total massive star formation timescale range from $\sim$ 1 to $5   \text{ times }10^5$~yr. The statistical lifetime of a phase is consequently obtained as the fraction of the total time equal to the number of objects in that phase with respect to the total number of objects. These methods depend on the assumption of a total time for the star formation process, however, that is constrained by simulations (e.g. \citealt{Davies11}) or derived with respect to the known age of OB stars (e.g. \citealt{Wood89, Motte07,Russeil10,Mottram11,Tige17}).\\
\indent An alternative way to quantify the timescales of the various steps is to study the effects that massive young stellar objects (YSOs) have on their immediate surroundings. The chemical evolution is affected by changes in density and temperature induced by the forming stars. During the so-called warm-up phase \citep[e.g.][]{Viti99, Viti01, Garrod08}, molecules formed on the surface of the dust grains are rapidly released into the gas phase, increasing their observed abundance by several orders of magnitude \citep{Viti04,Garrod13, Choudhury15}. Thus, the molecular composition of the material around YSOs contains information about the physical conditions of their past \citep{VanDerTak05}. A well-known example are the rich molecular spectral features produced by hot molecular cores (HMCs), which are compact regions around YSOs ($\lesssim 0.1$ pc radius; \citealt{Kurtz00,Cesaroni05}).\\
\indent Chemical tracers that show a relation between their observed abundances and the different phases of the star formation process are commonly called chemical clocks \citep[e.g.][as some recent examples; see also \citealt{vanDishoeck98} for a review]{Fontani07, Beuther09, Hoq13, Giannetti19, Urquhart19, Sabatini20}. 
Through the comparison of their observed abundances obtained from large samples of massive clumps at different evolutionary stages and those predicted by accurate time-dependent chemical models, it is possible to quantify the timescales of each stage. An attempt to adopt this type of analysis was carried out by \cite{Gerner14} based on the chemical properties of a sample of 59 massive clumps in different evolutionary stages, and modelling the warm-up phase alone. The radial profiles of temperature and density were assumed to be static with time and were described as power laws with the average properties of each evolutionary class (see their Sect.~5.1.1 for more details). The chemistry was then evolved in time until they found the best match with the average observed abundances. They interpreted this {time} as the typical age of each class. Based on these results, \cite{Gieser19} studied the chemical history of the HMC VLA 3 in the high-mass star-forming region AFGL 2591 to assess the effects of different initial chemical conditions on the model results and on the derived evolutionary timescales. They showed that the latter are sensitive to the model assumptions. More recently, \cite{Gieser21} used the same method to investigate 22 cores identified in 18 high-mass star-forming regions with ALMA.\\
\indent An alternative to these approaches is to adopt a time-dependent analytical function of the temperature to simulate the gradual warming-up of the clumps in a range of typical observed values \citep[i.e. $\sim$10-300 K; e.g.][]{Viti99, Garrod06}. \cite{Awad10} proposed an improved model of the warm-up phase  applied to the low-mass regime, where the temperature evolution was based on the radiative transfer (RT) calculation of \cite{Nomura04}. More recently, \cite{Bonfand19} described the evolution of the physical parameters of the high-mass star-forming region Sgr B2 with 3D-RT simulations. They followed the trajectory of a parcel of gas under free-fall collapse and modelled its chemistry by updating the temperature and density of the gas. In this paper, we have developed a model similar to \cite{Bonfand19} by employing a time-dependent description of the thermal evolution of massive clumps during the warm-up phase, with the aim to propose a new method for deriving the evolutionary timescales of the massive star formation process.\\ 
\indent This paper has the following structure: in Sect.~\ref{sec2:sample_tracers} we present the reference sample of massive clumps, the evolutionary sequence, and the chemical tracers we employed to derive the duration of each phase. The model and chemical network are described in Sect.~\ref{sec3:methods}. In Sect.~\ref{sec3:results} we present our results, the post-processing procedure, and the derivation of the durations of each evolutionary phase. In Sect.~\ref{sec4:discussion} we discuss our estimates and how the selected tracers are reproduced by the models, providing information on the reliability of the chemical clocks. Finally, in Sect.~\ref{sec5:conclusions} we summarise our conclusions.
\begin{figure*} 
\centering
        \includegraphics[width=0.83\textwidth]{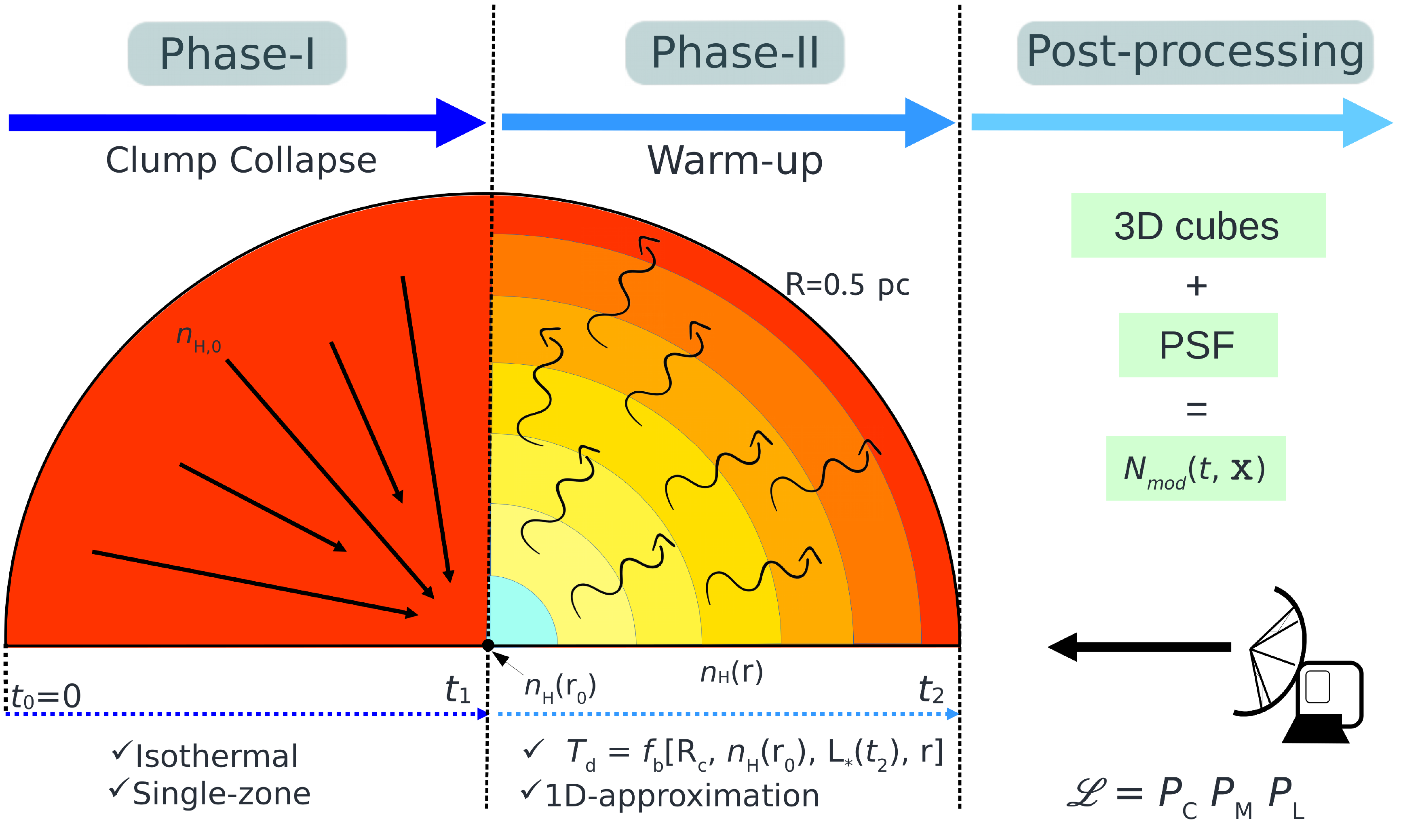}  
        \caption{Sketch of the physical model employed in this work. {\it Left panel:} Collapse phase (see Sect.~\ref{subsec:PH1_method}) solved in a single-zone approximation. The chemical output of phase I is used as initial condition of phase II. {\it Central panel:} Warm-up phase solved in a 1D approximation from 1 to 10$^5$ AU (see Sect.~\ref{subsec:PH2_method}). {\it Right panel:} Post-processing applied to compare the final chemical outputs of phase II with the column densities observed in the TOP100 (see Sections~\ref{sec43:column-densities} and~\ref{subsec41:comparison}). The physical parameter assumed in phases I and II are summarised in Tab.~\ref{Tab:par-space}. The time evolution of the model is indicated by the dotted blue arrows.}
        \label{fig:sketch}
\end{figure*}

\section{Reference sample and selected tracers}\label{sec2:sample_tracers}
\indent The survey we used to define the reference sample and the evolutionary stages of the massive star formation process is the APEX\footnote{{\it Atacama Pathfinder EXperiment 12 meter submillimeter telescope} \citep{Gusten06}} {\it Telescope Large Area Survey of the Galaxy} (ATLASGAL; \citealt{Schuller09}). ATLASGAL offers a complete view of the high-mass star-forming regions in the inner Galaxy at 870 $\mu$m and provides the ideal basis for detailed studies of a large number of massive clumps in different evolutionary stages, with the unprecedented angular resolution of $\sim$19\arcsec~ \citep{Contreras13, Csengeri14, Urquhart14a}. Estimates of luminosities, masses, kinematic distances, and dust temperature (\citealt{Urquhart14c}, \citealt{Wienen15} and \citealt{Urquhart18}) have been derived for more than $\sim$10$^4$ dust clumps.\\
\indent Using the unbiased nature of ATLASGAL, the ATLASGAL-TOP100 sample (hereafter TOP100) has been defined as a flux-limited sample of high-column density clumps, selected with additional infrared (IR) criteria to include sources that potentially cover the whole spectrum of ages \citep[see][]{Giannetti14}. The number of sources included in the TOP100 was slightly refined by {\cite{Konig17}}, and to date, it includes 111 clumps. Combining the IR properties of the massive clumps in the TOP100 sample with radio-continuum measurements at 5-9 GHz, the objects were divided into the following four evolutionary classes: (1) 70 $\mu${m} weak stage (70w; i.e. quiescent){,} constituted by sources undetected at 24 $\mu$m and showing no clear compact emission at 70 $\mu$m (or are seen in absorption at this wavelength). This stage represents the earliest phase of massive-star formation and potentially includes starless or prestellar cores. (2) Mid-IR weak stage (IRw; protostellar){,} composed of compact and visible sources at 70 $\mu$m that are {still} undetected at 24 $\mu$m or are associated with a weak IR source ($<$2.6 Jy fluxes). {The clumps in this evolutionary stage are young and} likely dominated by cold gas. (3) Mid-IR bright stage (IRb; high-mass YSOs){,} {made by the sources that show a strong mid-IR} emission at 8 and 24 $\mu$m, which was interpreted to be caused by the progressive dust heating induced by the forming (proto-)star(s). (4) HII regions (HII), {where the sources are} bright at 70 and 24 $\mu$m, and detected in radio continuum at 5-9 GHz.\\
\indent In the sample, the clump masses range from $\sim$18 to $\sim$5$\times$10$^4$ M$_\odot$. The bolometric luminosities are between $\sim$60 and $\sim$4$\times$10$^6$ L$_\odot$ and correspond to a range in luminosity-to-mass ratio ($L/M$) of $\sim$0.2-350 L$_\odot$/M$_\odot$. All the TOP100 clumps have the potential of forming high-mass stars and show no bias in terms of distance or mass between the evolutionary classes (\citealt{Konig17}).\\ 
\indent Four molecular tracers were studied with the APEX telescope in the TOP100 sample, revealing their potential as chemical clocks: formaldehyde \citep[\hhco;][]{Tang18}, methyl acetylene, acetonitrile, and methanol \citep[\chhhcch, \chhhcn~and \chhhoh, respectively;][]{Giannetti17_june}. The detection rates, abundances, and excitation temperatures derived from each molecular species increase with the $L/M$ of the clumps, which is known to trace the evolution of the star formation process \citep{Saraceno96, Molinari08, Urquhart18}. With respect to the line-of-sight (LOS) and beam-averaged column densities reported by \cite{Giannetti17_june} and \cite{Tang18}, we calibrated our method to estimate the duration of the four evolutionary phases defined in the TOP100 sample. We propose a general pipeline of comparisons of models and observations that can be expanded with additional tracers in future follow-ups.

\section{Methods}\label{sec3:methods}
In this section we present the model we developed to describe the time evolution of the abundances of the selected molecular tracers in the TOP100 survey (see Sect.~\ref{sec2:sample_tracers}). {We describe the physical and the chemical model separately in Sections~\ref{sec32:methods_phys_model} and~\ref{sec31:methods_chem_model}, respectively.}

\subsection{Physical model}\label{sec32:methods_phys_model}
Our physical model comprises two distinct stages that are sketched in Fig.~\ref{fig:sketch}. The first, namely the collapse (left panel of Fig.~\ref{fig:sketch}), describes the density evolution of an isothermal clump with a single-zone approximation, exploring different conditions as in \cite{Viti99} and \cite{Garrod06}, for example. In the second phase, the warm-up (central panel of Fig.~\ref{fig:sketch}), the gas and dust temperatures (assumed to be in equilibrium) evolve as a function of time, driven by the luminosity of the central forming protostar. The temperature profiles are computed with the radiative transfer code {\sc Mocassin} (see \citealt{Ercolano03, Ercolano05}), and the mass distribution of the clump is described by a static gas radial density profile (e.g. \citealt{Tafalla04}).\\
\indent In the following sections we describe the details of each phase. The specific details of the RT calculations will be presented in a forthcoming paper (\citealt{GrassiPREP}). 

\subsubsection{Phase I: Isothermal collapse}\label{subsec:PH1_method}
The first phase of the model simulates a semi-analytical single-zone isothermal collapse (see \citealt{Spitzer78, Brown88, Viti99}). The gas number density, $n_{\rm H}$, at the centre of the clump, evolves with time as
\begin{equation}\label{eq:modi_tff}
\frac{{\rm d}n_{\rm H}(t)}{{\rm d}t} = b \cdot \left(\frac{n_{\rm H}(t)^4}{n_{\rm H,0}}\right)^{1/3} \left\lbrace \mathcal{C}\:n_{\rm H,0} \left[\left(\frac{n_{\rm H}(t)}{n_{\rm H,0}}\right)^{1/3}- 1\right]\right\rbrace ^{1/2}\,,
\end{equation}
\noindent
where $n_{\rm H,0}$ is the initial central gas number density, $b$ is a factor which aims to mimic a slower collapse compared to the ideal free-fall time (i.e. $b=1$), $\mathcal{C}=24\pi G m_{\rm H}$, $G$ is the gravitational constant, and $m_{\rm H}$ is the hydrogen mass. Eq.~\ref{eq:modi_tff} is obtained assuming the conservation of mass during an isothermal contraction of a spherical clump (\citealt{Spitzer78}).\\
\indent We set the initial gas number density to the typical values of a clump, $n_{\rm H,0}={n(\rm{H}})+2n({\rm H}_2) = 10^{4}$ cm$^{-3}$ \citep{BerginTafalla07, Gerner14}. The collapse assumes a constant temperature of 15~K (for gas and dust temperatures; e.g. \citealt{Konig17}) and a visual extinction $A_{\rm v} = 10$ mag (e.g. \citealt{Semenov10, Reboussin14}). The cosmic-ray ionisation rate of hydrogen molecules was set to $\crir = 5 \times 10^{-17}$ s$^{-1}$, as observationally constrained by \cite{VanDerTak00} in high-mass star-forming regions, and in agreement with the results reported by \cite{Sabatini20}. The specific density of the dust grains was $\rho_0 = 3$ g cm$^{-3}$, typical of silicates (e.g. \citealt{Draine84}), the dust-to-gas ratio $\mathcal{D}=10^{-2}$, and we used a constant grain size $\left\langle a\right\rangle =0.1$ $\mu$m. The gas mean molecular weight was $\mu = 2.4$.\\
\indent To simulate various physical conditions for the environment in which the seeds of massive protostars are formed, the collapse was stopped at different final densities $n_{\rm H}(r_0)$ (see Fig.~\ref{fig:sketch}) that were observationally constrained (Tab.~\ref{Tab:par-space}; e.g. \citealt{Mueller02, Sabatini19}) and were then used as central densities during the warm-up phase.

\begin{table}
        \caption{Parameter space explored in our models; i.e. phase I (Sect.~\ref{subsec:PH1_method}) and phase II (Sect.~\ref{subsec:PH2_method}).}
        \setlength{\tabcolsep}{10pt}
    \renewcommand{\arraystretch}{1.3}
        \centering
        \begin{tabular}{clcc}
                \toprule
                \toprule
                &Parameter & Values & Unit \\
                \midrule
                (1)& $b$ & $1$; $0.5$; $0.1$ & $-$ \\           
                (2)& $n_{\rm H}(r_0)$ & $10^5$; $10^{6}$; $10^7$; $10^8$ & cm$^{-3}$ \\
                (3) & $R_{\rm c}$           & $10^4$; $10^{4.5}$; $10^5$ & AU \\
                (4) & $\dot{M}$ & $10^{-5}$; $10^{-3}$& ${\rm M}_\odot$ yr$^{-1}$ \\
                \bottomrule
                \bottomrule
        \end{tabular}\label{Tab:par-space}
\end{table}

\subsubsection{Phase II: Warm-up}\label{subsec:PH2_method}
The second phase of the model, sketched in the central panel of Fig.~\ref{fig:sketch}, simulates the warm-up induced by a protostar at the centre of a spherical clump by using a 1D approximation of 100 logarithmic radial steps from 1 to 10$^5$ AU (i.e. up to a typical clump size of $\sim$0.5 pc; e.g. \citealt{Motte18}).\\
\indent The mass distribution of the core is described by the gas radial density profile (e.g. \citealt{Tafalla02,Tafalla04}),
\begin{equation}\label{eq:rho-prof}
n_{\rm H}(r) =  n_{\rm H}(r_0) \frac{R_{\rm c}^{5/2}}{R_{\rm c}^{5/2}+r^{5/2}}\,,
\end{equation}
\noindent where $n_{\rm H}(r_0)$ is the central number density at $r_0$, at which phase I is stopped (see second row in Tab.~\ref{Tab:par-space}). This ensures continuity between the two physical phases of the model. In Eq.~\ref{eq:rho-prof}, $R_{\rm c}$ is the core radius, that is, the radius at which $n_{\rm H}(R_{\rm c})=0.5~n_{\rm H}(r_0)$. The slope 5/2 agrees with the observations in massive star-forming regions (e.g. \citealt{Mueller02, Schneider15}), while similar profiles were successfully applied to model the H$_2$ distribution in clumps and filamentary structures (e.g. \citealt{Beuther02,Arzoumanian11, Andre16, Sabatini19}). Alternatively, power-law profiles can be assumed to model the H$_2$ distribution especially on the clumps scale (e.g. \citealt{Gieser21}). This provides minor differences in the final clump mass (a factor of $\lesssim$2 smaller for a slope of 5/2).\\ 
\indent The physical parameters of {phase II} (i.e. $\crir$, $\mathcal{D}$, $\mu$, and $\left\langle a\right\rangle$) were assumed to be the same as those of the collapse phase, while the visual extinction at each radius $r$, was \mbox{$A_{\rm v} (r) = N({\rm H},r)/(2\times 10^{21}$ cm$^{-2}$)}, where $N({\rm H},r)$ is the column density of the hydrogen nuclei, obtained by integrating Eq.~\ref{eq:rho-prof} from the edge of the clump (e.g. \citealt{Tielens10,Zhu17}). We took the final abundances of {phase I} as initial chemical conditions for each radial grid point of {phase II}, rescaled by the density profile. The chemistry was evolved in time, assuming that the individual cells are independent throughout the entire time evolution.\\
\begin{figure} 
        \centering
        \includegraphics[width=1\columnwidth]{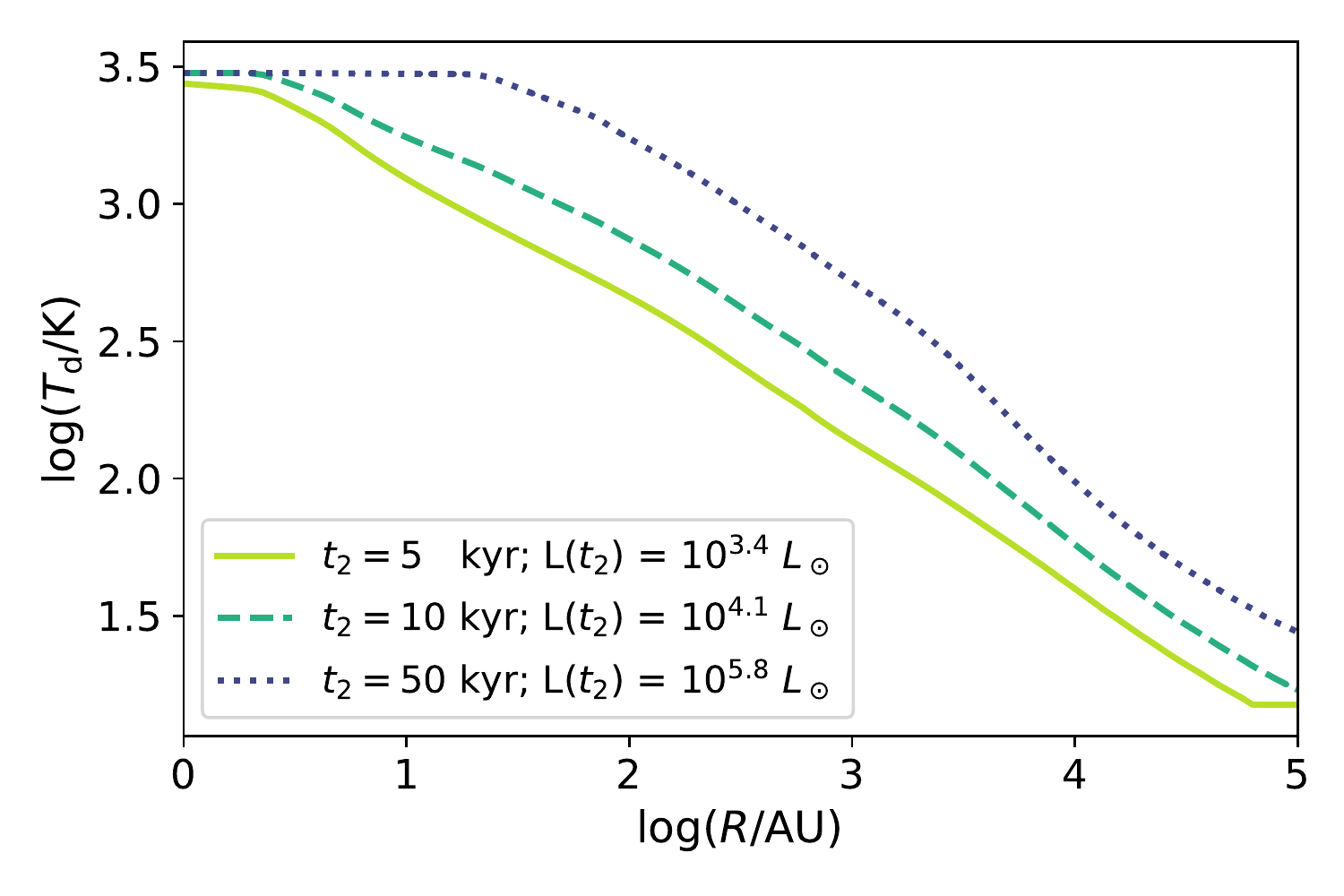}      
        \caption{Dust temperature radial profile at different times for an arbitrary model with the gas number density profile described by Eq.~\ref{eq:rho-prof} and $R_c = 10^{4.5}$ AU and $n_{\rm H}(r_0)=10^7$ cm$^{-3}$. With time, $T_{\rm d}$ increases, driven by the protostar luminosity. The flat inner part represents the sublimation of the dust grains. The minimum temperature is set to 15 K to ensure continuity with the collapse phase.}
        \label{fig:temp_prof}
\end{figure}
\indent To model the thermal evolution of the clump during the warm-up phase, we generated a set of models with different density distributions and protostar luminosities by employing the Monte Carlo radiative transfer code {\sc Mocassin}. The models provide an interpolable look-up table $T_{\rm d}$($r,t_2$) $= f_b $[$R_c$, $n_{\rm H}(r_0)$, $L_*(t_2)$, $r$], where $L_*(t_2)$ is the luminosity of the protostar and $t_2$ the time spent since the start of the warm-up phase. The function that describes how $L_*(t_2)$ evolves in time was derived by \cite{Hosokawa09} and depends on the mass accretion rate, $\dot M$, {assumed for the protostar}. Based on the very few observational estimates (e.g. \citealt{Beuther02-bis, Herpin12, Wang13}), we explored two typical cases: $\dot M=10^{-5}$~$\msun$~yr$^{-1}$ and $\dot M=10^{-3}$~$\msun$~yr$^{-1}$ (see Figures~4 and~12 in \citealt{Hosokawa09}). These values also agree with the results of \cite{Peters11}, who simulated the collapse of a magnetised rotating molecular cloud of 1000 M$_\odot$. During {phase II}, we assumed that the gas radial density profile of the clump and its dust content remained constant over time so that the mass accretion rate of the protostar was completely balanced by the average infall rate observed in a massive clump (e.g. \citealt{Schneider10, Peretto13, Wyrowski16,Liu18}).\\
\indent Analogously to {phase I}, we assumed that the gas and dust temperatures were in equilibrium; this assumption is less accurate at small radii (i.e. $\lesssim$50 AU), where the temperatures are close to the dust sublimation limit, and at large radii (i.e. $\sim$10$^5$ AU), where the gas-dust collision term is subdominant with respect to the radiation coupling (\citealt{Draine11}). However, neither region is relevant in our analysis and does not affect our findings. An example of how the $T_{\rm d}$ radial profile evolves is shown in Fig.~\ref{fig:temp_prof} for three (typical) times, $\dot M=10^{-3}$~$\msun$~yr$^{-1}$ , and for the following parameters $R_c = 10^{4.5}$ AU and $n_{\rm H}(r_0)=10^7$ cm$^{-3}$. For a qualitative comparison of the $T_{\rm d}$ radial profiles shown in Fig.~\ref{fig:temp_prof}, we refer to the recent results of \cite{Gieser21}, who derived the temperature structure of 18 massive star-forming regions based on two of the four tracers considered in this work (i.e. \hhco~and \chhhcn; Sect.~\ref{sec2:sample_tracers}). In this case, the authors assumed a power law (i.e. $T(r)\propto r^{-q}$; see their eq.~1) to fit the radial temperature in each source, providing an average slope of $|q| = 0.4 \pm 0.1$. Our profiles in Fig.~\ref{fig:temp_prof} show a similar slope ($|q|\sim0.5$) at distances larger than the dust sublimation radius (i.e. the flat region in the same figure) and comparable temperatures at their fiducial radius (see Tab.~3 in \citealt{Gieser21}). A more detailed description of the effect of $L_*(t_2)$ and $n_{\rm H}(r)$ on $T_{\rm d}$($r,t_2$) will be discussed in a forthcoming paper (\citealt{GrassiPREP}).\\
\indent To compare our results with the observed properties of the TOP100 clumps \citep{Giannetti14, Konig17}, we evolved one model for each mass in Tab.~\ref{Tab:MODELmasses} (derived from $n_{\rm H}(r_0)$ and $R_{\rm c}$ in Tab.~\ref{Tab:par-space}){, and for each value} of $b$ and $\dot{M}$ in Tab.~\ref{Tab:par-space}. We considered only clumps with a total mass higher than 15 M$_\odot$, which is the lowest mass associated with a TOP100 source, for a total of 54 models.

\begin{table}
	\caption{Summary of the total masses of each model.}
	\setlength{\tabcolsep}{10pt}
	\renewcommand{\arraystretch}{1.3}
	\centering
	\begin{tabular}{lccc}
		\toprule
		\toprule
		$ n_{\rm H}(r_0)$ &\multicolumn{3}{c}{$R_{\rm c}$/AU}\\
		\cmidrule(rl){2-4}
		[cm$^{-3}$] & $10^4$ & $10^{4.5}$ & $10^5$\\		
		\midrule
		
		$10^5$  & {\bf 0.4} M$_\odot$ & {\bf 3.8} M$_\odot$ & 18.5     M$_\odot$ \\
		$10^6$  & {\bf 3.5} M$_\odot$ &     38.4  M$_\odot$ & 184.6    M$_\odot$ \\
		$10^7$  &     34.7  M$_\odot$ &    383.7  M$_\odot$ & 1846.8   M$_\odot$ \\
		$10^8$  &    346.6  M$_\odot$ &   3836.7  M$_\odot$ & 18468.2  M$_\odot$ \\
		\bottomrule	
		\bottomrule
	\end{tabular}\label{Tab:MODELmasses}
	\tablefoot{{The models with masses lower than the lowest mass associated with the TOP100 sources \citep{Konig17} are marked in boldface and were excluded.}}
\end{table}

\subsection{Chemical model}\label{sec31:methods_chem_model}
\noindent
The chemical evolution of a massive clump (i.e. the collpase and the warm-up phases) is described by employing the publicly available time-dependent code \krome\footnote{\url{https://bitbucket.org/tgrassi/krome/wiki/Home}} \citep{Grassi14}.\\
\indent Adsorption and desorption\footnote{We include thermal and cosmic-ray induced desorption.} processes were included as in \cite{Hasegawa92} and \cite{Hasegawa93}. The surface reactions between two species $i$ and $j$ follow \cite{Semenov10}, where the rate coefficient, in units of cm$^3$ s$^{-1}$, is 
\begin{equation}\label{eq5:diff}
k_{ij}^s = \frac{P_{ij} (k_{\rm diff}^i + k_{\rm diff}^j)}{N_{\rm sites} n_{{\rm d}}}\,, 
\end{equation}
\noindent
with $k_{\rm diff}^i = \nu_0^i \exp (-T_{\rm diff}^i/T_{\rm d})$ the diffusion through thermal hopping ($T_{\rm diff}^i = 0.77 T_{\rm b}^i$; \citealt{Semenov10}), $T_{\rm d}$ is the dust temperature, $\nu_{0}^i = (2n_{\rm S}E_{\rm b}^i/\pi^{2}m_i)^{1/2}$ is the characteristic Debye vibration frequency for the adsorbed species, $n_{\rm S} = 1.5 \times 10^{15} $cm$^{-2}$ is the surface density of binding sites, $m_i$ is the mass of the species, and $E_{\rm b}^i = k_B T_{\rm b}^i$ is the binding energy of the $i$th species on the binding site. In Eq.~\ref{eq5:diff}, $n_{\rm d} = \mathcal{D} n_{\rm H} m_{\rm H} \mu/M_{\rm d}$ is the dust number density, where $M_{\rm d} = 4/3 \pi \rho_0 \left\langle a\right\rangle^3$ is the dust mass, and the total number of binding sites of a grain $N_{\rm sites} = 4\pi \left\langle a\right\rangle^2/a_{\rm pp}^2$ assumes an average distance between two contiguous sites of $a_{\rm pp} = 3$~\AA~(\citealt{Hocuk15}).\\
\indent The probability for a reaction to occur is $P_{ij} = \alpha_{ij} \exp (-E_a/k_B T_d)$, where $E_a$ is the activation energy of the reaction and $\alpha_{ij}$ is a parameter that depends on the number and on the type of species in the reaction. In the case of exothermic reactions (i.e. $E_a = 0$), (I) if $i \neq j$, $P_{ij} = 1$; (II) if $i = j$, $P_{ij} = \alpha_{ij} = 1/2$. For endothermic reactions (i.e. $E_a \neq 0$), $\alpha_{ij}$ is the inverse of the number of paths in the branching ratios.\\
\indent The final chemical network was derived from \cite{Semenov10}\footnote{Ohio StateUniversity (OSU) chemical network, version March 2008 (\url{http://www.mpia.de/homes/semenov/Chemistry_benchmark/model.html})} and contains 654 chemical species and 5869 gas-phase, gas-grain, and grain-surface reactions. The photochemical rate coefficients follow the $A_{\rm v}$ formalism as in \cite{Draine78} (see KIDA\footnote{\url{http://kida.astrophy.u-bordeaux.fr/}}). In order to ensure that each depleted species was released into the gas phase, we modified the original network by adding 70 missing desorption processes, with the binding energy values updated to the most recent estimates in KIDA (see Tab.~\ref{Tab:number_reac} and Tab.~\ref{Tab:bindings} for the reactions, and Appendix~\ref{app:benchmark} for the network benchmark).\\
\indent The abundances of chemical species evolve with time, starting from the assumed initial conditions following the recent large-scale simulations of molecular clouds (e.g. \citealt{Hocuk16, Clark19}), showing that CO is already formed at densities of ${\rm few}\times 10^3$ cm$^{-3}$. Following these findings, we assumed H, C, and O to be in molecular form. In particular, the abundances of H$_2$, H$_3^+$, He, N, O, CO, and N$_2$ were taken from \cite{Bovino19}, while for the other elements, we refer to \cite{Garrod06}, as reported in Table~\ref{Tab:ICC_PH1}. 

\begin{table}
	\caption{Summary of fiducial initial elemental abundances, $n_i$,
		with respect to the abundance of H-nuclei, $n_{\rm H}$.}
	\setlength{\tabcolsep}{10pt}
	\renewcommand{\arraystretch}{1.2}
	\centering
	\begin{tabular}{lc|lc}
		\toprule
		\toprule
		species 	  	&  ($n_i/n_{\rm H}$)$_{\rm t=0}$ & species 	  	&  ($n_i/n_{\rm H}$)$_{\rm t=0}$ 	\\
		\midrule
		
		H$_2$ 	& 5.00(-1)	& Si 	  & 1.95(-6) \\
		He   	& 1.00(-1)	& S    	  & 1.50(-6) \\
		O 		& 1.36(-4)  & Fe      & 7.40(-7) \\
		CO 	    & 1.20(-4)  & P 	  & 2.30(-8) \\
		N 		& 1.05(-5)  & Na 	  & 2.00(-8) \\
		N$_2$	& 5.25(-6)  & Cl 	  & 1.40(-8) \\
		Mg 	    & 2.55(-6)  & H$_3^+$ & 3.18(-9) \\
		\bottomrule	
		\bottomrule
	\end{tabular}\label{Tab:ICC_PH1}
\tablefoot{{The ``A(-B)'' notation assumed in the table means ``A$\times$10$^{\rm -B}$''.}}
\end{table}

\begin{figure*} 
\centering
        \includegraphics[width=0.77\textwidth]{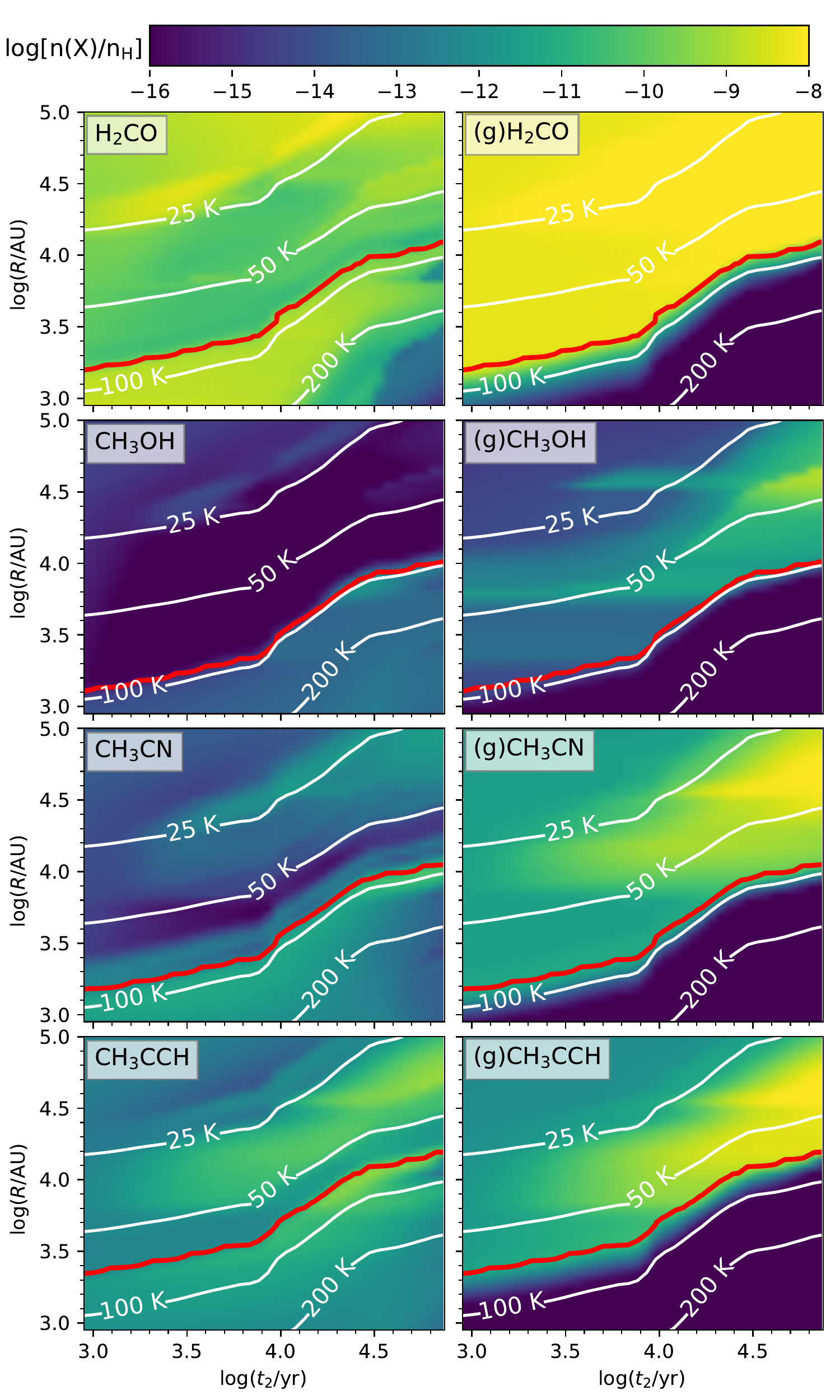} 
        \caption{Temporal and radial evolution of the abundances of the tracers observed in the TOP100 \citep{Giannetti17_june, Tang18} during {phase II} in the gas phase ({\it left panels}) and on dust ({\it right panels}) for the same reference model as in Fig.~\ref{fig:temp_prof}. White contours indicate the temperature computed with {\sc Mocassin}. Red curves corresponds to the evaporation front of the given tracer.}
        \label{fig:Wup1}
\end{figure*}

\begin{figure*} 
\centering
        \includegraphics[width=0.85\textwidth]{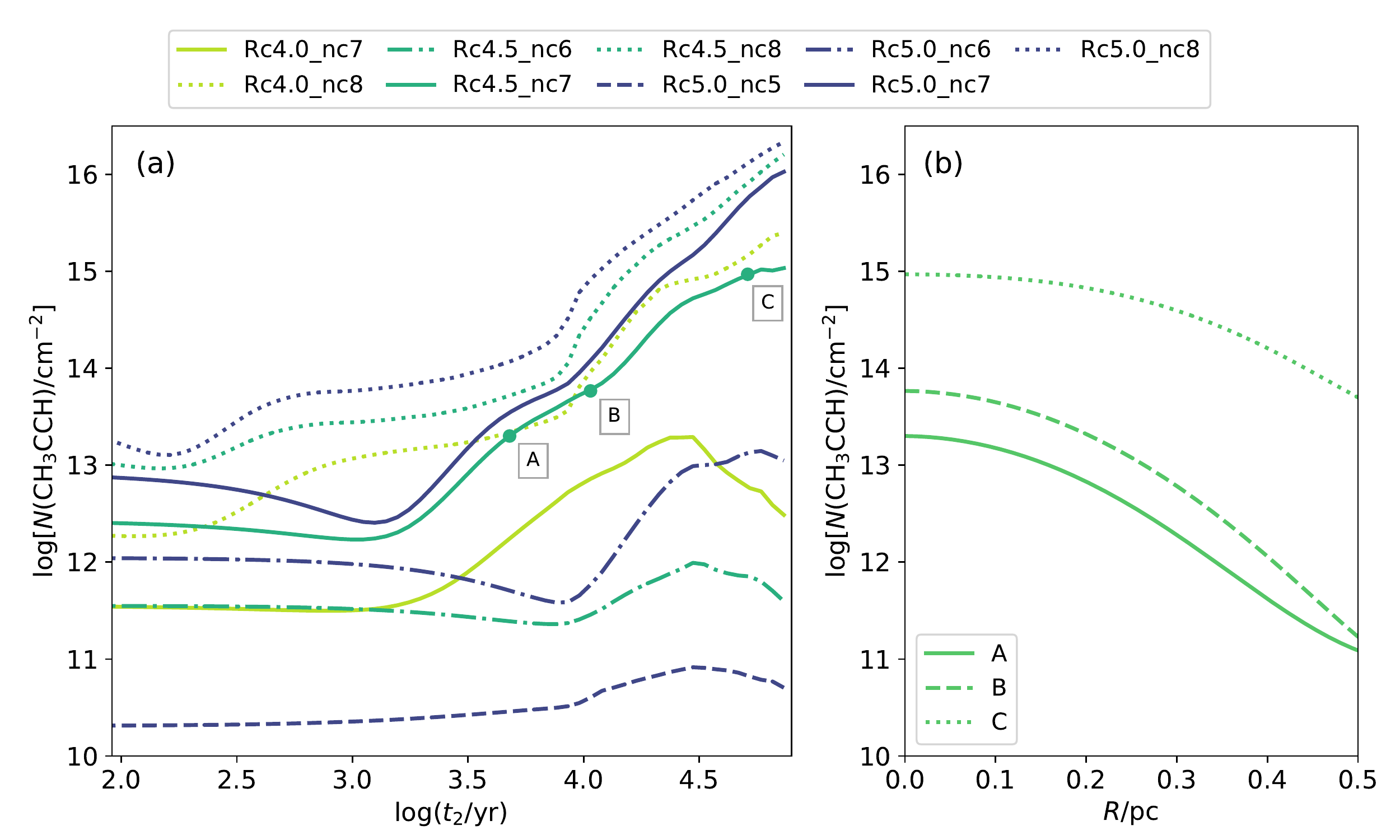}            
        \caption{{\it Panel} ($a$): Example of the LOS gas-phase column density profiles obtained at the end of the post-processing described in Sect.~\ref{sec43:column-densities} for models with $b=1$ and $\dot M = 10^{-3} M_\odot$ yr$^{-1}$ and for an arbitrarily chosen source at a distance of 3.5 kpc. For the {\tt U$\_$W} items in the legend, {\tt U} and {\tt W} are $R_c$ (colours) and $n_{\rm H} (r_0)$ (styles), respectively. {\it Panel} ($b$): Radial distribution of the gas-phase column densities extracted at the positions A, B, and C in panel ($a$), i.e. the same times as in Fig.~\ref{fig:temp_prof}.}
        \label{fig:smoothed_CH3CCH}
\end{figure*}

\section{Results}\label{sec3:results}
An example of the chemical evolution of some observed species during {phase I} is reported in Appendix~\ref{app:PH1}, while here we focus on {phase II} and the comparison with observations. However, because the outcome of {phase I} represents the initial chemical conditions for {phase II}, a few considerations for {phase I} should be made. The fractional abundances are considerably affected by the dynamics of the collapse, showing a difference up to $\sim$3 orders of magnitudes between the fastest ($b=1$) and the slowest ($b=0.1$) collapse, even if the global trend remains generally unaltered (more details in Appendix~\ref{app:PH1}). Based on the continuity between {phase I} and {phase II}, the final effect that $b$ has on the models is to provide different inputs for the warm-up phase. For this reason, we ran {phase II} for each of the 54 models defined by the parameter space, taking the effect of parameter $b$ on the second phase of the model into account (see also Sect.~\ref{subsec:PH2_method}).\\
\indent Fig.~\ref{fig:Wup1} shows the abundances of \hhco, \chhhcch, \chhhcn,~and \chhhoh~as a function of time and distance in the gas phase (left panels) and on grains (right panels; labelled `(g)') as a result of {phase II}. The white contours show the temperature, and the red curves the \textup{}evaporation fronts, namely the distance from the protostar at which the desorption timescale of a given species (defined as the inverse of the desorption rate, involving the thermal and the cosmic-ray induced desorption; \citealt{Semenov10}) is shorter than the dynamical timescale of the model. As a consequence, at distances larger than the evaporation front, the abundances of the tracers on dust grains (Fig.~\ref{fig:Wup1}) increase rapidly. At the same time, the temperatures at these distances are low enough to enhance two-body reactions on the surface of dust grains. Hydrogenation chains play a major role in the formation of \chhhcch, \chhhcn, and \hhco~on dust grains, and methanol is also formed through the reaction ${\rm (g)CH_3} + {\rm (g)OH} \rightarrow {\rm (g)CH_3OH}$.\\
\indent Each evaporation front grows as a function of time following the temperature evolution and ranges between $\sim$10$^3$ and $\sim$10$^5$ AU. Our findings agree with those of \cite{Choudhury15}, where the efficiency of the evaporation as a function of the thermal evolution of the clump is responsible for evaporation scales similar to ours (see their Fig.~B1).

\subsection{Post-processing of the model outputs} \label{sec43:column-densities}
The chemical structure of each clump was reconstructed assuming that each radial profile from {phase II} (i.e. each column in the relative abundance maps in Fig.~\ref{fig:Wup1}) represents a radius of a spherically symmetric clump. For each profile we generated a data cube of 150$\times$150$\times$150 pixels at a resolution of $\sim$10$^3$ AU/pixel, which corresponds to one-half of the size of the smallest evaporation front radius in Fig.~\ref{fig:Wup1}, in order to always sample the region at which the gas-phase abundances are enhanced by the desorption of the products of dust-phase chemistry. The column density maps of each tracer were then obtained by integrating the abundances along the LOS, applying for each tracer a convolution to the APEX resolution of the observed transitions\footnote{This step was done using the {\tt fft2} function of {\tt numpy.fft}: \url{https://numpy.org/doc/stable/reference/generated/numpy.fft.fft2.html}} (see Appendix~\ref{app:Nobs}), and taking the different distances of the TOP100 objects into account. The final column density profile of a given tracer \X, $N_{\rm mod}$(\X), was calculated along the LOS at the centre of the clump.\\
\indent With respect to this procedure, \chhhoh~and \chhhcn~required an additional constraint to reproduce the observed higher-K lines. In their spectral fit, \cite{Giannetti17_june} distinguished the hot and cold components, assuming two domains separated at 100 K and providing the individual column densities. Following the same approach provides additional constraints to reduce the uncertainties and to test the reliability of the temperature of the background model, especially in the innermost part of the clump. After reconstructing the spherical distribution of \chhhcn~and \chhhoh, we applied a 100 K temperature-mask, taking only regions into account in which $T>100$ K, to generate the averaged column densities profiles of the hot components alone.\\
\indent As an example, we report the final gas-phase \chhhcch~profiles in Fig.~\ref{fig:smoothed_CH3CCH}$a$, obtained assuming a distance of 3.5 kpc (arbitrarily chosen) to apply the post-processing. We show the results for $\dot M=10^{-3}$~$\msun$~yr$^{-1}$ and for the standard free-fall collapse model (i.e. $b=1$), and we note that $n_{\rm H}(r_0)$ is the parameter that mostly affects the evolution of $N_{\rm mod}$(\chhhcch).\\
\indent This behaviour is the consequence of two combined effects.
First, the initial abundances of the warm-up phase are scaled by the density profile of each model, and thus, models with higher values of $n_{\rm H}(r_0)$ have higher initial column densities (see Fig.~\ref{fig:collapseprofs}). Second, the different thermal structures of the clumps have a strong effect on the chemical evolution. At a given time, the average temperature along the LOS is lower in clumps with larger $n_{\rm H}(r_0)$ because the radiation of the protostar is more attenuated. Because the luminosity of the protostar increases with time and affects the temperature of the surrounding gas, the evaporation front grows and so does the number of chemical species released into the gas. In Fig.~\ref{fig:smoothed_CH3CCH}$b$ we show the radial profile of gas-phase $N$(\chhhcch) at three times (A, B, and C in panel a). The column density of this species increases by about one order of magnitude over time, caused by the expanding evaporation front and by the decrease in \chhhcch~adsorption rate in the outer parts of the model. The adsorption process is directly proportional to the density, which  drops at $r>R_{\rm c}$ (Fig.~\ref{fig:Wup1}). This also provides an explanation for the observed increasing trend in the gas-phase \chhhcch~abundance with the evolutionary stage of massive clumps (e.g. \citealt{Molinari16, Giannetti17_june}).

\subsection{Comparison of the modelled column densities with sources from the dTOP100 sample}\label{subsec41:comparison}

\begin{figure} 
        \includegraphics[width=1\columnwidth]{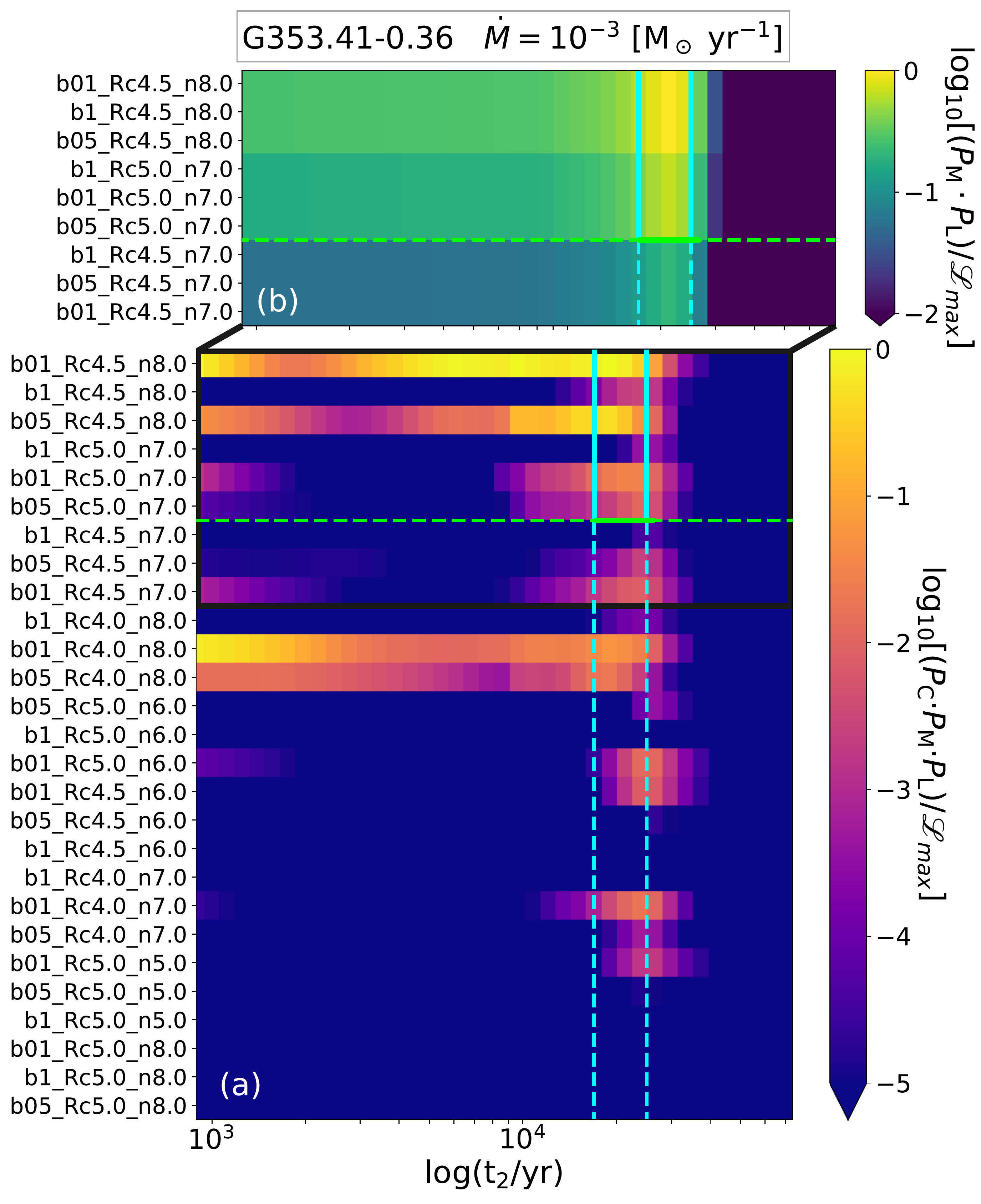}      
        \caption{$Panel\:(a)$: Example of the $\LL$ matrix calculated for G353.41-0.36 in all the models with $\dot M = 10^{-3} M_\odot$ yr$^{-1}$. Models on the y-axis are sorted from the top to the bottom by the difference between the mass of the models (see Tab.~\ref{Tab:MODELmasses}) and the value associated in the TOP100 (see Tab.~\ref{Tab:observed-prop}). For the sake of clarity, the colour bar is limited to five orders of magnitude and is normalised to maximum likelihood ($\mathscr{L}_{\rm max}$). $Panel\:(b)$: Zoom on the age limits identified by $\ppl$ and $\ppm$ showing how the likelihood would appear if $\ppc$ in Eq.~\ref{eq:lik} were not taken into account. Cyan vertical lines indicate the range of time that corresponds to the uncertainties on the observed luminosity in the TOP100, and the horizontal green line shows the same limit associated with the clump mass. Times and $\LL$ are shown in log-scale.}
        \label{fig:PROBmodVSobs}
\end{figure}

To infer the ages of the clumps and the relative duration of the evolutionary phases in the TOP100 sample, we compare the column densities derived from the observations (Appendix~\ref{app:Nobs} and Tab.~\ref{Tab:observed-prop}) with those from our post-processed models. For a given clump the dataset, $\DD= \{L, M, N({\tt X})\}$, comprises eight (two+six) measurements, that is, the luminosity $L$, the mass $M$, and the column densities of the six observed chemical tracers, $N({\tt X})$. At each time step of the warm-up phase, with the free parameters $\tte=\{b, \dot M, R_{\rm c}, n_{\rm H}(r_0)\}$ given the data $\DD$, the likelihood is
\begin{equation}\label{eq:lik}
\LL(\tte|\DD) = \ppc(\tte|\DD)\:\ppm(\tte|\DD)\:\ppl(\tte|\DD)\,,
\end{equation}
where $\ppc(\tte|\DD)$, $\ppm(\tte|\DD),$ and $\ppl(\tte|\DD)$ are probability density distributions of the chemical abundances, the mass, and the bolometric luminosity of each clump, respectively.\\
\indent The first factor in Eq.~\ref{eq:lik} is
\begin{equation}\label{eq:totchemprob}
\ppc = \prod_{\tt X} \ppcx\,,
\end{equation}
where for the sake of simplicity, we omitted the arguments of $\ppc$, and where \citep{Garrod07,Gerner14}
\begin{equation}\label{eq:chem-prob}
\ppcx = {\rm erfc}\left\lbrace \frac{|{\rm log}_{10}[N_{\rm obs}({\tt X})]-{\rm log}_{10}[N_{\rm mod}({\tt X})]|}{\sqrt{2}\sigma}\right\rbrace,
\end{equation}
with `erfc' the complementary error function. In Eq.~\ref{eq:chem-prob}, $N_{\rm obs}({\tt X})$ and $N_{\rm mod}({\tt X})$ are the observed and modelled column densities, respectively. As in \cite{Garrod07}, we set $\sigma = 1$, which corresponds to a difference of one order of magnitude between the observed and the modelled column densities. This assumption takes into account the uncertainties on the observed column densities, on the initial chemical conditions assumed for the {phase I}, and on the shape of the density profile assumed in {phase II} (see Sect.~\ref{subsec:PH2_method}). For sources without a detection, we assumed a detection limit (see Appendix~\ref{app:Nobs}) and set $\ppcx = 1$ when $N_{\rm mod}$({\tt X}) $< N_{\rm obs}$({\tt X}), while we used Eq.~\ref{eq:chem-prob} when $N_{\rm mod}$({\tt X}) $> N_{\rm obs}$({\tt X}).\\
\indent For each clump, $\ppl$ and $\ppm$ are Gaussian functions with mean equal to the $L$ and $M$ values derived by \cite{Konig17}. The standard deviations were computed as follows: We added in quadrature a 50\% uncertainty associated with the models based on the approximation of a single clump size of 0.5 pc and a constant accretion rate over time (Sect.~\ref{subsec:PH2_method}), in addition to the errors of 50\% on $L$ and 20\% on $M$ proposed by \cite{Urquhart18}. This gives a standard deviation for $\ppl$ of about 60\% of $L$, and for $\ppm$ of about 50\% of $M$. We note that an additional variation of $10\%$ on these uncertainties does not produce significant variations in our final results.\\
\indent Fig.~\ref{fig:PROBmodVSobs}a shows an example of how $\LL$ varies as a function of time when models with the same mass accretion rate ($\dot{M}=10^{-3}\,\msun$~yr$^{-1}$) are considered. The models are sorted in ascending order by the difference in mass between the values reported in \cite{Konig17} and those of the models (i.e. Tab.~\ref{Tab:MODELmasses}), while $t_2$ is the time spent in the warm-up phase defined in Sect.~\ref{subsec:PH2_method}.
In both panels, $\ppl$ defines the range of time spent by the protostar in the warm-up phase (i.e. the range of time in between the vertical cyan lines), which only depends on the uncertainties associated with the bolometric luminosity of each source. Analogously, $\ppm$ is constrained by the mass range found for a given source (e.g. see discussions in \citealt{Konig17} and \citealt{Urquhart18}), and it is therefore possible to define a lower mass limit that is indicated by the horizontal green line. The combined information given by $\ppl$ and $\ppm$ limits the age range of each source. In this area, the models with the same density profile defined by Eq.~\ref{eq:rho-prof}, but different values of $b$, are degenerate, showing the same $\LL$ as a function of $t_2$ (see Fig.~\ref{fig:PROBmodVSobs}b). $\ppc$ is fundamental to break this degeneracy (see Fig.~\ref{fig:PROBmodVSobs}a), providing further information on the dynamics of the collapse that led to the observed chemical properties of each clump, and placing additional constraints on determining the relative duration of the evolutionary classes in the TOP100.\\
\indent The absolute time is hence the sum of the collapse timescales, $t_1$, and the time spent during the warm-up phase, $t_2$. The collapse time depends on the setup of {phase I}, and it ranges between $(0.4$-$5)\times 10^6$ yr depending on $b$ and $n_{\rm H}(r_0)$, while the warm-up time depends on the mass accretion rate of {phase II} (see Sect.~\ref{subsec:PH2_method}).\\
\indent The absolute age for the $\ell$th source is the weighted mean,
\begin{equation}\label{eq:tA}
t_{{\rm A},\ell}= \left[\sum_{i,j} \frac{{\rm log}_{10}[t_{1,i,\ell}+t_{2,j,\ell}]}{\left({\rm log}_{10}\LL_{i,j,\ell}\right)^2}\right] \left[\sum_{i,j} \frac{1}{\left({\rm log}_{10}\LL_{i,j,\ell}\right)^{2}}\right]^{-1}\,,
\end{equation}
\noindent where $i$ ranges over the 54 models, $j$ over the 100 times used to sample {phase II}, and we used the log-likelihood as weight.\\
\indent We performed the comparison described above only for the sources in the TOP100 sample that agreed within the uncertainties with the masses of the models in Tab.~\ref{Tab:MODELmasses}. This reduced the sample to 48 objects, that is, 6 sources in the 70w stage, 16 IRw, 15 IRb, and 11 HII regions. Nevertheless, this limitation does not affect our approach because our model employs a subset of objects that are representative of the whole population, preserving their global features (e.g. \citealt{Csengeri16, Giannetti17_june, Konig17}). We summarise the observed properties of the clumps in this subsample in Tab.~\ref{Tab:observed-prop}. It is worth noting that while the TOP100 sample does not show bias in terms of the source distances for each evolutionary phase (e.g. \citealt{Konig17}), in Tab.~\ref{Tab:observed-prop} only the IRw class extends beyond $\sim$10 kpc. To verify whether the distance has an effect on our results, we performed a test considering sources up to a maximum distance of 6 kpc. We did not find significant changes in the final durations reported in Sect.~\ref{sec4:discussion}. 

\section{Discussion}\label{sec4:discussion}
In this section, we estimate the relative duration of the evolutionary phases identified in the TOP100, {following a different approach compared to} the statistical lifetimes. Our findings are then compared with the relative number of objects observed in ATLASGAL (\citealt{Urquhart18}), and we discuss whether the synthetic column densities of the selected chemical tracers match the observed densities.

\subsection{Estimates of the duration of evolutionary phases}\label{sec:duration_estimates}
To compare the ages of the evolutionary stages defined in the TOP100 with the average absolute time estimated in a sample of $\mathcal{N}$ objects, $\left\langle \tA\right\rangle = \mathcal{N}^{-1}\sum_{\rm n=1}^{\mathcal{N}}t_{{\rm A}, n}$, we defined the age factor of the $\ell$th source as
\begin{figure} 
        \includegraphics[width=1\columnwidth]{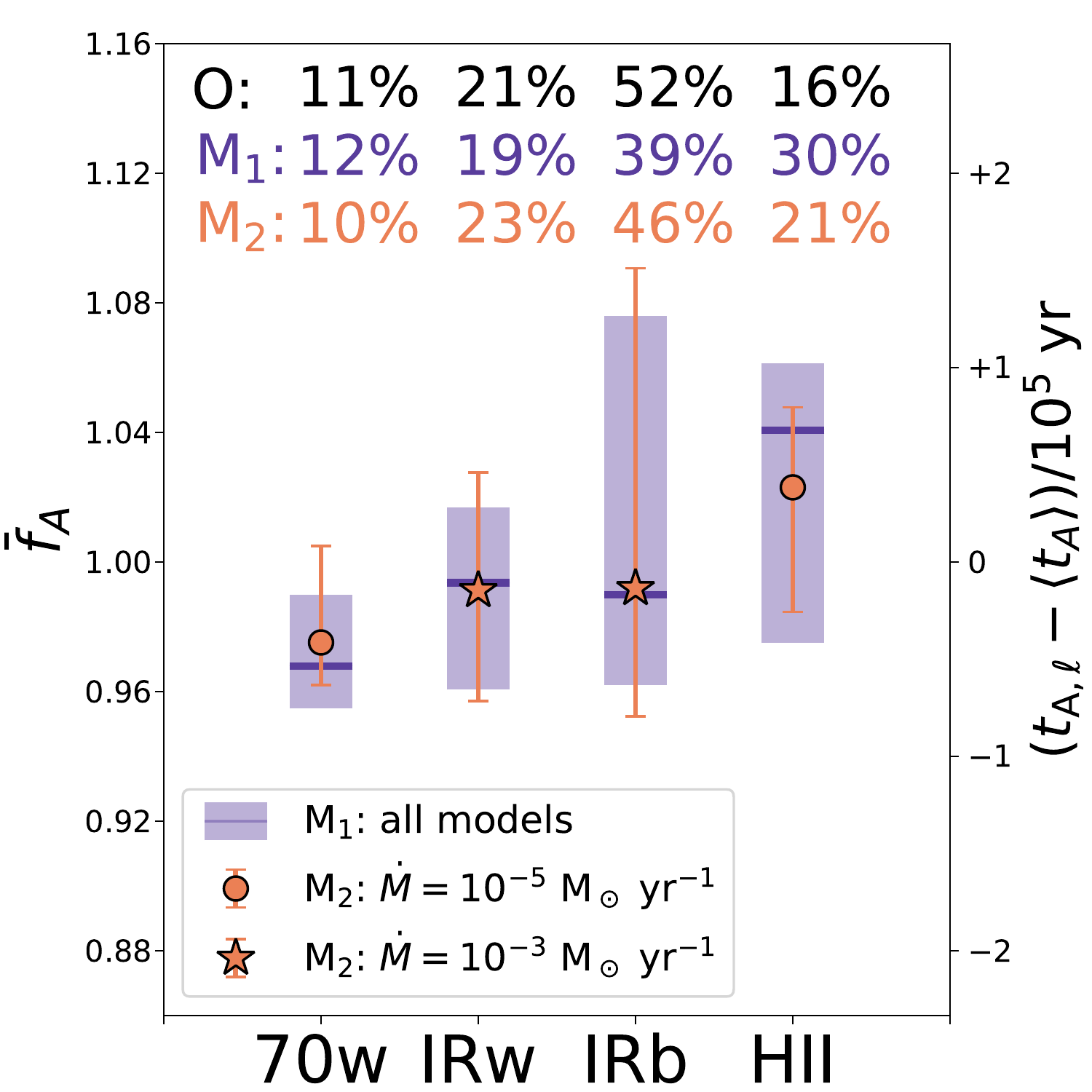}
        \caption{Summary of the final durations estimated in each evolutionary class defined for the TOP100. The relative number of massive clumps in each phase and observed in ATLASGAL are shown in black \citep[``O''; ][]{Urquhart18}. Purple shaded areas indicate the 5th and 95th percentiles of the age factor, $\fA$, and the corresponding value in yr in our sample (model ``M$_1$''), and purple markers (`--') represent their medians. The numbers associated with the model ``M$_2$'' (orange), indicate the same quantities obtained by separating the models with respect to $\dot M$ before calculating the average age of each source (see text). The different markers in M$_2$ show the combination of $\dot{M}$ that best match the observed relative number of object in each phase (legend).}
        \label{fig:rel_duration}
\end{figure}
\begin{equation}\label{eq:fA}
f_{\rm A, \ell} = \frac{t_{\rm A, \ell}}{\left\langle \tA\right\rangle}\,.
\end{equation}
In this context, the models provide $\left\langle \tA\right\rangle \sim 10^6$ yr starting from $t_1=0$ (i.e. the beginning of {phase I}; see also Fig.~\ref{fig:sketch}). The average $\fA$ were estimated as the median of the age-factor distributions in each evolutionary class, computed with Eq.~\ref{eq:fA} and reported as purple markers in Fig.~\ref{fig:rel_duration}. For the sake of clarity, Fig.~\ref{fig:rel_duration} shows the corresponding value of $t_{\rm A, \ell} - \left\langle \tA\right\rangle$ yr that is associated with each $\fA$. {We quantified the duration of an evolutionary phase, $\Dtp$,} as the time between the minimum and maximum value of $t_{\rm A, \ell} - \left\langle \tA\right\rangle$ (i.e. the lower and upper limits of the purple shaded areas in Fig.~\ref{fig:rel_duration}). These values were computed as the $5$th and $95$th percentiles of the {\bf $\fA$} distributions in each evolutionary phase. The sum of the four $\Dtp$ is the total time of the high-mass star formation process $\tmsf$. The purple percent values in Fig.~\ref{fig:rel_duration} (M$_1$) indicate the contribution of each phase to the total time $\tmsf$.\\
\indent We find that following the classification of \cite{Giannetti14} and \cite{Konig17}, 12\% of the star formation time is spent in the early phase (70w), and the IRw stage is associated with 19\% of $\tmsf$. This suggests a fast evolution during the early stage of the massive star formation process, which in total correspond to the $\sim$30\% of $\tmsf$. Advanced stages (i.e. IRb sources) have the longest duration (39\%), while the remaining 30\% of the $\tmsf$ is spent in the final stage (HII).\\
\indent Assuming that the number of objects in an evolutionary stage is also representative of its duration, we compared our findings with the relative number of objects per evolutionary class in ATLASGAL. This is represented as black percentages in Fig.~\ref{fig:rel_duration} (``O''; from \citealt{Urquhart18}).
The duration found for the early stages agrees with the observed classification, and the later stages are probably biased by the different definition of IRb and HII in the TOP100 and ATLASGAL. In particular, when radio continuum emission is found at either 4 or 8 GHz within 10\arcsec~of the ATLASGAL peak, the source is classified as HII in the TOP100 (e.g. \citealt{Konig17}). Different criteria have been applied in \cite{Urquhart18} to classify the advanced stages that contain radio bright HII regions, massive young stellar objects, and sources associated with methanol masers. These surveys have a different coverage than the complete ATLASGAL and also have different sensitivities (see \citealt{Urquhart14c} for more details), so that the final number of objects in the advanced stages might be underestimated. Because this limit affects the separation between IRb and HII, we note that if we consider a single phase to describe the more evolved sources (i.e. IRb + HII), the total duration associated with this phase would agree better.\\
\indent An additional source of uncertainty is a different value of $\dot{M}$ throughout the evolutionary sequence. Evidence of an increasing mass accretion rate in the intermediate stages of the massive star formation process is discussed in \cite{Beuther02-bis}, but when the protostars are close to the main sequence, their radiation pressure might slow down or quench the mass accretion \citep[e.g.][]{Nakano95, StahlerPalla04, Klassen12}. 
To quantify the effects of $\dot M$, we repeated the calculation of the duration of each phase by separating the models by accretion rate and mixing the results of the different phases in all their possible combinations. We find that different $\dot M$ produces different durations and $\tmsf$. The most accurate solution to interpret the observations in ATLASGAL shows an accretion rate that is initially slow and increases during the two intermediate classes, to decrease again in the more evolved phases (see the orange symbols and percentages in Fig.~\ref{fig:rel_duration}; M$_2$). The latter result allows us to verify how reliable the assumption of a constant H$_2$ radial density profile is that we made for the clumps during {phase II} (see Sect.~\ref{subsec:PH2_method}). In the most variable scenario where $\dot M=10^{-3}$~$\msun$~yr$^{-1}$ over the $t_2$ identified in our models (and assuming no mass replenishment), the majority of the clumps shows a mass loss of $\lesssim$20\%. Only two clumps of $\sim$20~$\msun$ would suffer a relevant mass loss (i.e. G353.07+0.45 and G316.64-0.09 in Tab.~\ref{Tab:observed-prop}). However, these two sources belong to the IRb and IRw phases, which contain the majority of the clumps, and can be considered outliers in these classes. Conversely, for sources in the 70w and HII evolutionary phases, where $\dot M=10^{-5}$~$\msun$~yr$^{-1}$, the mass loss is negligible. Alongside the few observational estimates available for $\dot M$ and for the infall rate (e.g. from the parent filament; see Sect.~\ref{subsec:PH2_method}), these results mean that the assumption of a constant-density H$_2$ profile is reasonable on a clump scale.\\
\begin{figure*} 
\centering
        \includegraphics[width=0.93\textwidth]{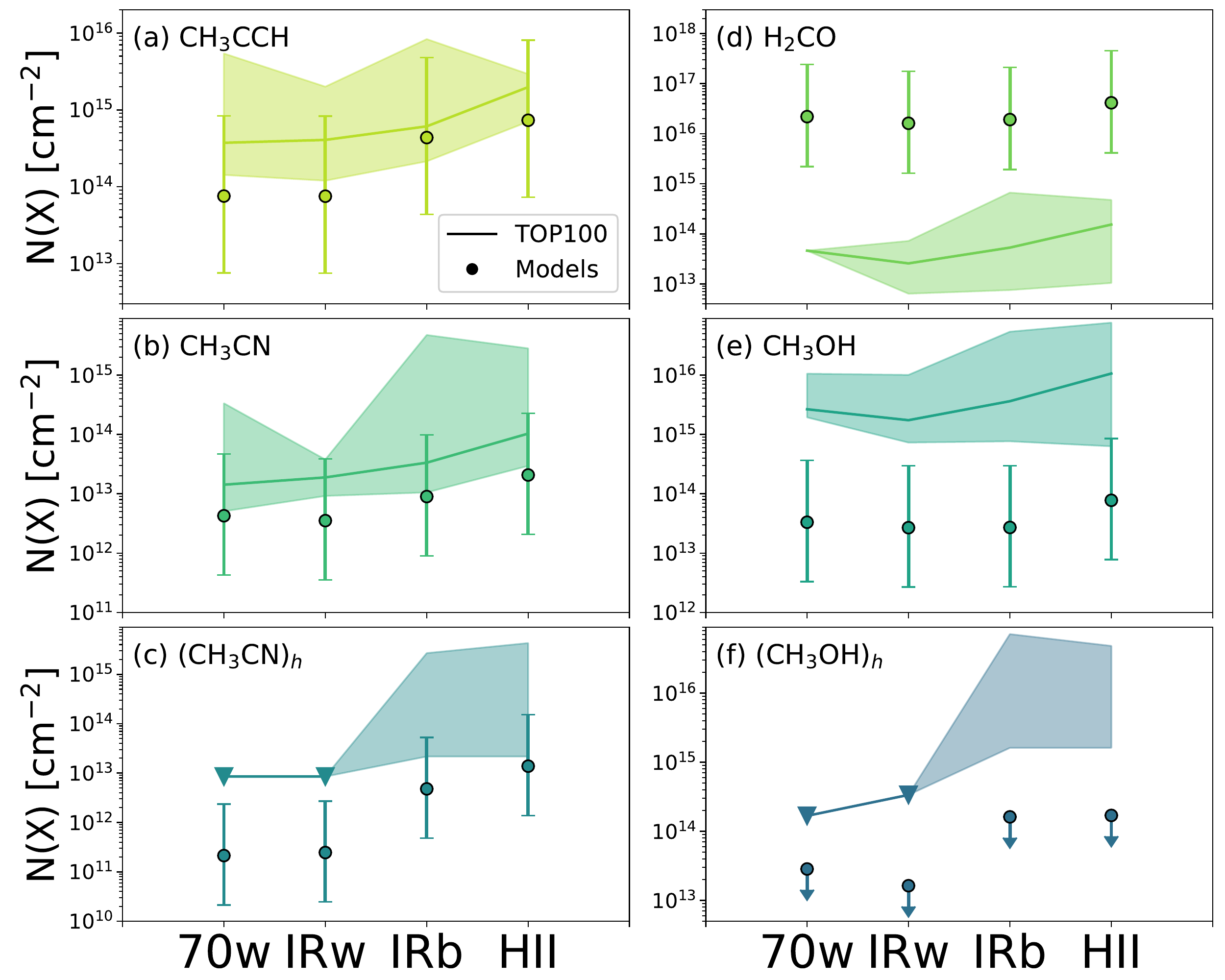}
        \caption{Comparison of the median column densities observed in the TOP100 (solid lines) with those predicted by our models (circles) that were obtained by applying the same procedure as was used to quantify $\Dtp$ (see Sect.~\ref{subsec41:comparison}). Panels are separated by tracers (colours). The shape of the shaded areas indicates the minimum and maximum observed column density, and the error bars associated with each circle incorporate the uncertainty of one order of magnitude assumed in Eq.~\ref{eq:chem-prob} for the comparison. The modelled column densities of the hot component of methanol (circles in panel f) are multiplied by a factor 1000. Triangles indicate the upper limits. The column densities are shown in log-scale.}
        \label{fig:final-abb}
\end{figure*}
\indent The individual durations estimated in M$_2$ lead to $\sim$5~$\times~10^4$ yr for younger objects (i.e. 70w), $\sim$1.$2\times10^5$ yr for the IR-weak, $\sim$2.$4\times10^5$ yr for IR-bright sources, and $\sim$1.$1\times10^5$ yr for HII regions. Additionally, the average $\fA$ of each class increases with the evolution of the sources (circles and stars in Fig.~\ref{fig:rel_duration}), so that the typical object of each class appears statistically older than the average from the previous evolutionary phase. We therefore assume that M$_2$ is the most representative model, with a total massive star formation time $t_{\rm MSF, M_2}\sim5.2\times10^{5}$ yr ($t_{\rm MSF, M_1} \sim$1.1 $t_{\rm MSF, M_2}$).

In this context, the chemistry plays a crucial role in breaking the model degeneracy when the likelihood is defined by $\ppm$ and $\ppl$ alone and in matching the results from ATLASGAL (see Fig.~\ref{fig:rel_duration}). If $\ppc$ is not included in $\LL$ (Eq.~\ref{eq:lik}), the duration of the advanced stages shows a difference up to a factor of $\sim$8 when compared to those predicted by M$_2$. This corresponds to a total time,  $\tmsf$, 4.5 times shorter, and the final relative durations differ from the observed relative number of objects in each class by up to a factor of 5.

\subsection{Selected tracers and chemical clocks}
Four molecules were selected in this work: formaldehyde, methyl acetylene, acetonitrile, and methanol (see Sect.~\ref{sec2:sample_tracers}). They all manifest an observed upward trend in their abundances with increasing evolutionary stage, which is indicated by the luminosity-to-mass ratio of the clumps \citep{Giannetti17_june,Tang18}. Two thermal components were also required to reproduce the observed emission of the higher K-ladders of methanol and acetonitrile lines in \cite{Giannetti17_june}.\\
\indent To evaluate to which extend the models are able to reproduce the observed column densities, the same procedure as we used to calculate $\Delta t_{\rm phase}$ was employed (Equations~\ref{eq:tA} and~\ref{eq:fA}). Figure~\ref{fig:final-abb} summarises the comparison of the modelled and observed column densities, grouped by tracer type (colours) and evolutionary class. Shaded areas represent the range of the observed beam-averaged column densities (Sect.~\ref{sec2:sample_tracers} and Appendix~\ref{app:Nobs}), with an order of magnitude uncertainty (Eq.~\ref{eq:chem-prob}), not shown in the figure. The median values of these observed column densities are shown as a solid line. The median column densities predicted by our models are plotted as coloured circles and are shown with an uncertainty of one order of magnitude, the same as assumed for the observations.

\subsubsection{Methyl acetylene and acetonitrile}
The reliability of \chhhcch~and \chhhcn~as thermometers has been widely discussed in the past (e.g. \citealt{Zhang98, Molinari16, Giannetti17_june}). In particular, \cite{Giannetti17_june} detected an increasing trend of the temperature and column density of \chhhcch~from the less to the more evolved sources in the TOP100 (i.e. shaded area in panel a of Fig.~\ref{fig:final-abb}). The same behaviour is found for \chhhcn~if the contributions of a hot and cold component are taken into account.\\
\indent In Fig.~\ref{fig:final-abb} the observed column densities of methyl acetylene (panel a) and acetonitrile (panels b, and c) derived from the observations increase by a factor of 10 between the 70w and the HII stages. The column densities provided by our models reflect the same trends and reproduce the observed values. The discrepancy between models and observations is within the uncertainties.\\
\indent The robustness of \chhhcch~ and \chhhcn~as chemical clocks was finally tested by removing these two tracers from $\ppc$. This had the same effect as removing $\ppc$ from $\LL$ altogether (see Sect.~\ref{sec:duration_estimates}), and we conclude that methyl acetylene and acetonitrile are effective chemical clocks for characterising the evolution of massive star-forming clumps and that they contribute to constraining our findings.

\subsubsection{Formaldehyde and methanol}
Unlike \chhhcch~and \chhhcn, the model column densities of methanol and formaldehyde (panels d, e, and f in Fig.~\ref{fig:final-abb}) agree less well with the observed densities. The slightly increasing trend in the observed column densities with evolution is reproduced by the models, but the modelled column densities under- or over-estimate those observed for the TOP100 by one order of magnitude at least, similarly to \cite{Gerner14}. This might be related to the uncertainties in the chemical pathways that determine the formation of methanol and its precursors. For temperatures below 20 K, hydrogenation chains are usually invoked to convert CO into \hhco~and \chhhoh~on grains because hydrogenation is made very efficient by the strong CO depletion that occurs at low temperatures (e.g. \citealt{Caselli08}, \citealt{Giannetti14} and \citealt{Sabatini19}). However, at the same temperatures, thermal desorption is inhibited and methanol is not efficiently released from the dust into the gas phase, suggesting that the surface chemistry alone is not capable of explaining the detection of gas-phase methanol in the cold phases of massive star-forming regions as well (e.g. \citealt{Cosentino18}). Alternative formation paths and mechanisms have been proposed to solve this issue (e.g. \citealt{Viti99,Garrod07,Vasyunin13}), but this was found to be relatively inefficient in reproducing the observed abundances of methanol \citep{Geppert06}, and to no longer accurately predict the abundances of other chemical species \citep{Garrod07}.
The same issue also concerns \hhco, a molecular precursor of methanol. In the advanced evolutionary stages, the amount of methanol is underestimated (although closer to the observed values). The reason might be that the thermal evaporation of methanol in the warmer environments is not well determined, or that further formation pathways are missing in the colder phases. We also note that the absence of the quantum tunneling diffusion in our models might affect the final abundance of methanol, enhancing the efficiency of CO hydrogenation in the colder evolutionary phases, and thus influencing the final abundance of methanol in the warm phases \citep{Vasyunin17}.\\
\indent An additional uncertainty that we explored is the effect of different activation energies for reactions that drive the formation of molecules such as formaldehyde and methanol at low temperatures. Recently, a study of CO hydrogenation on water ice by \cite{Rimola14} reviewed the activation energy of the hydrogenation reaction between \hhco~and H producing CH$_3$O, the precursor of methanol. The authors proposed a barrier of 1300 K, which would favour the formation of \chhhoh~compared to the value commonly employed ($\sim$2500 K, as in this work and derived for the gas phase; \citealt{Woon02}). Even a lower activation energy of $\sim$500 K has been predicted by \cite{Fuchs09}. However, we have tested these two scenarios without finding significant changes with respect to the modelled column densities of the tracers in Fig.~\ref{fig:final-abb}. The most substantial change concerns \chhhoh~when the extreme activation energy proposed by \cite{Fuchs09} is assumed. In this case, methanol column densities approach the values observed in the two most advanced evolutionary stages, while those of formaldehyde remain always largely overestimated. This suggests that the activation energy has secondary effects in reproducing the observed abundances of methanol and formaldehyde, and other processes or observational strategies, such as different or multiple transitions, should be taken into account.\\
\indent We finally note that the discrepancies in the abundances of \hhco~and \chhhoh~are not relevant for our results. When these two tracers are removed from $\ppc$ and $\LL$, the duration of each phase and the final $t_{\rm MSF}$ remain unaffected, showing a variation of $\sim$10\% with respect to the numbers reported in Fig.~\ref{fig:rel_duration}.

\section{Summary and conclusions}\label{sec5:conclusions}
We have presented a new and generalised method for deriving the evolutionary timescales of the massive star formation process. Compared to typical statistical approaches (e.g. \citealt{Russeil10, Mottram11, Tige17, Motte18, Urquhart18}), our method follows a different procedure in which the $\tmsf$ is the final result that is derived as the sum of the individual $\Dtp$ (Sect.~\ref{sec:duration_estimates} and Fig.~\ref{fig:rel_duration}), and not an a priori assumption.\\  
\indent We developed a set of 54 models, built to represent the entire population of massive clumps of the ATLASGAL-TOP100 sample (Sect.~\ref{sec1:intro}). The models consist of two physical phases: an initial isothermal collapse followed by a warm-up phase induced by a massive protostar at the centre of a spherical clump (Sect.~\ref{sec3:methods}). In particular, we assumed a Plummer-like density profile with a slope of 5/2, while the temperature profiles were derived from accurate 3D RT simulations. {The recent empirical estimates of the temperature profiles obtained by \citealt{Gieser21} from the same tracers used in this work agree with our computed profiles (see Sect.~\ref{subsec:PH2_method})}. We then compared the column densities of formaldehyde (\hhco), methyl acetylene (\chhhcch), acetonitrile (\chhhcn), and methanol (\chhhoh) as observed in the ATLASGAL-TOP100 sources \citep{Giannetti17_june, Tang18} with the modelled densities by post-processing the outputs of the models at the same angular resolution as the observed data (see Appendix~\ref{app:Nobs}).\\ 
\indent The timescales of the evolutionary stages associated with the massive star formation process were derived by considering the physical properties of the clumps, that is, their mass and luminosity, and the observed abundances of each selected molecular tracer. Considering a mass accretion rate depending on the evolutionary phases, we find a total star formation time, $\tmsf \sim 5.2 \times 10^5$ yr, which agrees well with that assumed by the statistical methods. This provides a new and significant validation of the statistical methods. The individual $\Dtp$ that define $\tmsf$ are found to be $\sim$5~$\times~10^4$ yr for the 70w, $\sim$1.$2\times10^5$ yr for the IRw, \mbox{$\sim$2.$4\times10^5$ yr} for IRb sources, and $\sim$1.$1\times10^5$ yr for HII evolutionary stages of the TOP100 sample (Sect.~\ref{sec2:sample_tracers}). Knowing $\tmsf$ and $\Dtp$, we derived the relative duration of each phase and found an agreement with the relative number of objects in each phase observed in the ATLASGAL survey. This provides further confirmation of the reliability of the method we presented. The chemical constraint included in the likelihood to determine the duration of the different phases is necessary to achieve results that agree with the observed relative number of objects in the ATLASGAL survey. Without this additional constraint, $\Dtp$ becomes shorter by up to a factor of $\sim$8 with respect to those reported above, and the relative durations differ by up to a factor of 5 with the relative number of objects in each class.\\
\indent Of the selected molecular tracers, \chhhcch~and \chhhcn~are best reproduced by our models, and \hhco~and \chhhoh~differ from the observed values, although they do follow the observed trends. Therefore the final $\Dtp$ reported in this paper are mainly constrained by \chhhcch~and \chhhcn. Because the chemical constraint is important to identify reliable timescales, we plan to extend the number and complexity of the selected molecular tracers with several chemical clocks proposed in the literature (e.g. HC$_3$N by \citealt{Taniguchi18}; CH$_3$OCHO and CH$_3$OCH$_3$ by \citealt{Coletta20}; see also \citealt{Urquhart19, Belloche20}).
This might help to better assess the different phases of the massive star formation process and increase the number of constraints on $\Dtp$ reported here.\\
\indent We also found that the evaporation fronts of the discussed molecular tracers vary between 10$^3$ and 10$^5$ AU during the warm-up phase (Sect.~\ref{sec3:results}), which in our models marks the regions in which each tracer becomes abundant in the gas phase. This result suggests that we compare our findings with the observations provided by the new astronomical facilities. In particular, the {\it Atacama Large Millimeter/submillimeter Array} (ALMA; \citealt{Wootten09}) offers the perfect opportunity of achieving the high angular resolutions (e.g. \citealt{Csengeri18,Maud19,Sanna19,Johnston20}) needed to sample scales of about the physical sizes of the modelled evaporation fronts. Moreover, ALMA can also reach the sensitivities needed to detect a large number of components in the $J_{\rm {K_a, K_c}}$ band. This would allow the removal of possible opacity effects and the accurate definition of the thermal state of the clump that can be compared with the results of this work. The increasing complexity of chemical models and the progress made in the observational techniques make this paramount goal increasingly achievable in the near future.\\
\indent To conclude, we have reported relevant and robust results on the high-mass star formation process together with reliable estimates of the duration of the different phases of this complex process. 
The present pipeline is based on 1D models, which cannot capture the effect of dynamical processes such as magnetic fields and turbulence on the density evolution of the collapsing clumps and the subsequent (proto-)stellar accretion process fully. On the other hand, low-dimensionality models allow distinguishing between the different chemical processes, and building a large number of models 
to infer statistical properties that can be compared with observations. Although some authors have recently included the  evolution of large chemical networks in 3D models \citep[see e.g.][]{Bovino19}, in the presence of a (proto-)stellar object, modelling the chemistry and the thermal evolution of the gas in 3D still represents a challenge.

\begin{acknowledgements}
{\rm The authors wish to thank the anonymous referee, for her/his suggestions to improve the manuscript, D. Semenov for fruitful discussions and feedbacks about the chemical model benchmark, and J. Ramsey how developed the \krome~python-interface employed in this work (see also \citealt{Gressel20}). This paper is based on data acquired with the Atacama Pathfinder EXperiment (APEX). APEX is a collaboration between the Max Planck Institute for Radioastronomy, the European Southern Observatory, and the Onsala Space Observatory. This research made use of Astropy, a community-developed core Python package for Astronomy (\citealt{Astropy13, Astropy18}; see also \url{http://www.astropy.org}), of NASA’s Astrophysics Data System Bibliographic Services (ADS) and of Matplotlib (\citealt{Matplotlib07}). GS acknowledges R. Pascale for useful suggestions and comments, and the {\it University of Bologna} which partially provided the funds ({\it Marco Polo fellowship}) for this project. SB acknowledges for funds through BASAL Centro de Astrofisica y Tecnologias Afines (CATA) AFB-17002.}
\end{acknowledgements}

%
%

\bibliographystyle{aa} 
\bibliography{mybib_GAL}

\begin{appendix}
\section{Details of the chemical network}\label{app:chem_det}
Tab.~\ref{Tab:number_reac} shows the types of chemical reactions that we included in the network. Column (1) contains the names of each type of reaction; column (2) an example of chemical reaction between two generic reactants A and B, and, column (3) summarises the total number of reactions included in the network.

\begin{table}
\caption{Summary of the reactions in the chemical network.}
\setlength{\tabcolsep}{3pt}
\renewcommand{\arraystretch}{1.3}
\centering
\footnotesize
\begin{tabular}{l|cc}
 \hline
 \hline
 Reaction	  	&  Example\tablefootmark{a}   & $\#$included\tablefootmark{b}  	\\
 \hline
 
 	Recomb. on grain 		& A$^{+}+$ grain$^{-}$ $\rightarrow$ A + grain		  			        & $13$\\
 	CR ionisation			& AB + $cr$ $\rightarrow$ AB$^{+}$ + e$^{-}$				  		    & \rdelim\}{2}{10.5mm}[\parbox{0.5mm}{\:\:220}]\\
 	CR photodissociation	& AB + $cr$ $\rightarrow$ A + B     		  		   				    &  \\
 	Gas-phase reactions		& (see the note below)\tablefootnote{The gas-phase processes included are ion-molecule, neutral-neutral, charge exchange, radiative association, radiative recombination and dissociative recombination.}                     & $4016$ \\
 	Photo-ionisation		& AB + $\gamma_{\rm UV}$ $\rightarrow$ AB + e$^{-}$  				        & \rdelim\}{2}{10.5mm}[\parbox{0.5mm}{\:\:153}]\\
 	Photo-dissociation		& AB + $\gamma_{\rm UV}$ $\rightarrow$ A + B         				        &   \\
 	Grain-surf. reactions	& A$_{\rm dust}$ + B$_{\rm dust}$  $\rightarrow$ AB$_{\rm dust}$ 				    & $266$ \\
 	Thermal desorpion		& AB$_{\rm dust}$ + $heat$ $\rightarrow$ AB	 					    & $195$ \\
 	Desorp. induced by CR	& AB$_{\rm dust}$ + $cr$ $\rightarrow$ AB  			  					    & $195$ \\
 	(CR photodiss.)$_{\rm dust}$	& AB$_{\rm dust}$ + $cr$ $\rightarrow$ A$_{\rm dust}$ + B$_{\rm dust}$	  		    & $185$ \\
 	(CR photoion.)$_{\rm dust}$	& AB$_{\rm dust}$ + $cr$ $\rightarrow$ AB$^{+}_{\rm dust}$ + e$^{-}$  		    & $56$ \\
 	(FUV photodiss.)$_{\rm dust}$	& AB$_{\rm dust}$ + $\gamma_{\rm UV}$ $\rightarrow$ A$_{\rm dust}$ + B$_{\rm dust}$ & $204$ \\
 	(FUV photoion.)$_{\rm dust}$	& AB$_{\rm dust}$ + $\gamma_{\rm UV}$ $\rightarrow$ AB$^{+}_{\rm dust}$ + e$^{-}$   & $171$\\
 	(Freeze-out)$_{\rm dust}$	& AB $\rightarrow$ AB$_{\rm dust}$ + $\gamma$	                            & $195$ \\
 \hline	
 \hline
\end{tabular}\label{Tab:number_reac}
\tablefoot{{\tablefoottext{a}{A and B are two generic reactants.}
\tablefoottext{b}{number of reactions.}}}
\end{table}

\noindent In Tab.~\ref{Tab:bindings} we provide the complete list of binding energies assumed in this work, updated to the most recent estimates found in KIDA. Where it was not possible to find a recent estimate, we used \cite{Semenov10} (species in bold). The 35 chemical species in the bottom part of the table are those added to the original network due to the missing desorption processes (see Sect.~\ref{sec31:methods_chem_model}).

\begin{table}
\caption{Complete list of binding energies.}
\setlength{\tabcolsep}{1pt}
\renewcommand{\arraystretch}{1.2}

\small
\centering
\begin{tabular}{rr|rr|rr|rr}
\toprule
\toprule
\multicolumn{8}{c}{Chemical species - T$_b$/K}\\
\cmidrule(rl){1-8}
C               &  10000  &  C$_6$H$_6$     &  7590  &  CO                 &    1300   &  HNC$_3$       &     4580 \\
C$_{10}$        &   8000  &  C$_7$          &  5600  &  CO$_2$             &    2600   &  HNCO          &     4400 \\
C$_2$           &  10000  &  C$_7$H         &  6140  &  CS                 &    3200   &  HNO           &     3000 \\
C$_2$H          &   3000  &  C$_7$H$_2$     &  6590  &  Fe                 &    4200   &  HS            &     2700 \\
C$_2$H$_2$      &   2590  &  C$_7$H$_3$     &  7040  &  FeH                &    4650   &  HSS           &     2650 \\
C$_2$H$_3$      &   2800  &  CH$_3$C$_6$H   &  7490  &  H                  &     650   &  Mg            &     5300 \\
C$_2$H$_4$      &   2500  &  C$_7$N         &  6400  &  H$_2$              &     440   &  MgH           &     5750 \\
C$_2$H$_5$      &   3100  &  C$_8$          &  6400  &  {\bf H$_2$C$_3$N}  &{\bf 5030} &  MgH$_2$       &     6200 \\
CH$_3$CH$_2$OH  &   5400  &  C$_8$H         &  6940  &  {\bf H$_2$C$_3$O}  &{\bf 3650} &  N             &      720 \\
C$_2$H$_6$      &   1600  &  C$_8$H$_2$     &  7390  &  {\bf H$_2$C$_5$N}  &{\bf 6630} &  N$_2$         &     1100 \\
C$_2$N          &   2400  &  C$_8$H$_3$     &  7840  &  {\bf H$_2$C$_7$N}  &{\bf 8230} &{\bf N$_2$H$_2$}&{\bf 4760}\\
CCO             &   1950  &  C$_8$H$_4$     &  8290  &  {\bf H$_2$C$_9$N}  &{\bf 9830} &  Na            &    11800 \\
C$_2$S          &   2700  &  C$_9$          &  7200  &  H$_2$CN            &    2400   &  NaH           &    12300 \\
C$_3$           &   2500  &  C$_9$H         &  7740  &  H$_2$CO            &    4500   &  NaOH          &    14700 \\
C$_3$H          &   4000  &  C$_9$H$_2$     &  8190  &  H$_2$CS            &    4400   &  NH            &     2600 \\
C$_3$H$_2$      &   3390  &  C$_9$H$_3$     &  8640  &  H$_2$O             &    5600   &  NH$_2$        &     3200 \\
C$_3$H$_3$      &   3300  &  C$_9$H$_4$     &  9090  &  H$_2$O$_2$         &    6000   &  NH$_2$CHO     &     6300 \\
C$_3$H$_3$N     &   5480  &  C$_9$N         &  8000  &  H$_2$S             &    2700   &{\bf NH$_2$OH}  &{\bf 6810}\\
CH$_3$CCH       &   3800  &  CH             &   925  &  HSSH               &    3100   &  NH$_3$        &     5500 \\
C$_3$N          &   3200  &  CH$_2$         &  1400  &  {\bf H$_3$C$_5$N}  &{\bf  7080}&  NO            &     1600 \\
C$_3$O          &   2750  &  H$_2$CCN       &  4230  &  {\bf H$_3$C$_7$N}  &{\bf  8680}&  NS            &     1900 \\
C$_3$S          &   3500  &  H$_2$CCO       &  2800  &  {\bf H$_3$C$_9$N}  &{\bf 10300}&  O             &     1600 \\
C$_4$           &   3200  &  CH$_3$N        &  5530  &  {\bf H$_4$C$_3$N}  &{\bf  5930}&  O$_2$         &     1200 \\
C$_4$H          &   3740  &  CH$_2$NH$_2$   &  5530  &  {\bf H$_5$C$_3$N}  &{\bf  6380}&  O$_2$H        &     5000 \\
C$_4$H$_2$      &   4190  &  CH$_3$O        &  4400  &  HCCNC              &    4580   &  O$_3$         &     2100 \\
C$_4$H$_3$      &   4640  &  CH$_3$         &  1600  &  HC$_2$O            &    2400   &  OCN           &     2400 \\
C$_4$H$_4$      &   5090  &  CH$_3$C$_3$N   &  6480  &  HC$_3$N            &    4580   &  OCS           &     2400 \\
C$_4$N          &   4000  &  CH$_3$C$_4$H   &  5890  &  {\bf HC$_3$O}      &{\bf 3200} &  OH            &     4600 \\
C$_4$S          &   4300  &  CH$_3$C$_5$N   &  7880  &  HC$_5$N            &    6180   &  S             &     2600 \\
C$_5$           &   4000  &  CH$_3$C$_6$H   &  7490  &  HC$_7$N            &    7780   &  S$_2$         &     2200 \\
C$_5$H          &   4540  &  CH$_3$C$_7$N   &  9480  &  HC$_9$N            &    9380   &  Si            &    11600 \\
C$_5$H$_2$      &   4990  &  CH$_3$CHO      &  5400  &  {\bf HCCN}         &{\bf 3780} &  SiC           &     3500 \\
C$_5$H$_3$      &   5440  &  CH$_3$CN       &  4680  &  HCN                &    3700   &  SiH           &    13000 \\
CH$_3$C$_4$H    &   5890  &  CH$_2$NH$_2$   &  5530  &  HNCCC              &    4580   &  SiH$_2$       &     3600 \\
C$_5$N          &   4800  &  CH$_3$OCH$_3$  &  3150  &  HCO                &    2400   &  SiH$_3$       &     4050 \\
C$_6$           &   4800  &  CH$_3$OH       &  5000  &  HCOOCH$_3$         &    6300   &  SiH$_4$       &     4500 \\
C$_6$H          &   5340  &  CH$_4$         &  9600  &  CH$_2$O$_2$        &    5570   &  SiO           &     3500 \\
C$_6$H$_2$      &   5790  &  CH$_3$NH$_2$   &  6580  &  HCS                &    2900   &  SiS           &     3800 \\
C$_6$H$_3$      &   6240  &  H$_2$CN        &  2400  &  He                 &     100   &  SO            &     2800 \\
C$_6$H$_4$      &   6690  &  CN             &  2800  &  HNC                &    3800   &  SO$_2$        &     3400 \\
\midrule
\multicolumn{8}{c}{Added in this work}\\
\midrule
CH$_3$COCH$_3$  &   3500  &  H$_2$SiO       &  4050  &  NO$_2$             &    2400   &  SiC$_3$       &     5100 \\
C$_3$P          &   5900  &  HCCP           &  4750  &  P                  &    1100   &  SiC$_3$H      &     5550 \\
C$_4$P          &   7500  &  HCL            &  5170  &  PH                 &    1550   &  SiC$_4$       &     5900 \\
CCl             &   1900  &  HCP            &  2350  &  PH$_2$             &    2000   &  SiCH$_2$      &     3750 \\
CCP             &   4300  &  HCSi           &  3630  &  PN                 &    1900   &  SiCH$_3$      &     4200 \\
CH$_2$PH        &   2600  &  HNSi           &  5080  &  PO                 &    1900   &  SiN           &     3500 \\
Cl              &   3000  &  HPO            &  2350  &  SiC$_2$            &    4300   &  SiNC          &     4300 \\
ClO             &   1900  &  N$_2$O         &  2400  &  SiC$_2$H           &    4700   &  SiO$_2$       &     4300 \\
CP              &   1900  &  NH$_2$CN       &  5560  &  SiC$_2$H$_2$       &    5200   &                &          \\
\bottomrule
\bottomrule
\end{tabular}\label{Tab:bindings}
\tablefoot{{Binding energies were taken from \cite{Wakelam17} and  \citet{Semenov10} (see text).}}
\end{table}

\section{Benchmark of the chemical network}\label{app:benchmark}

\begin{figure*} 
\includegraphics[width=0.9\textwidth]{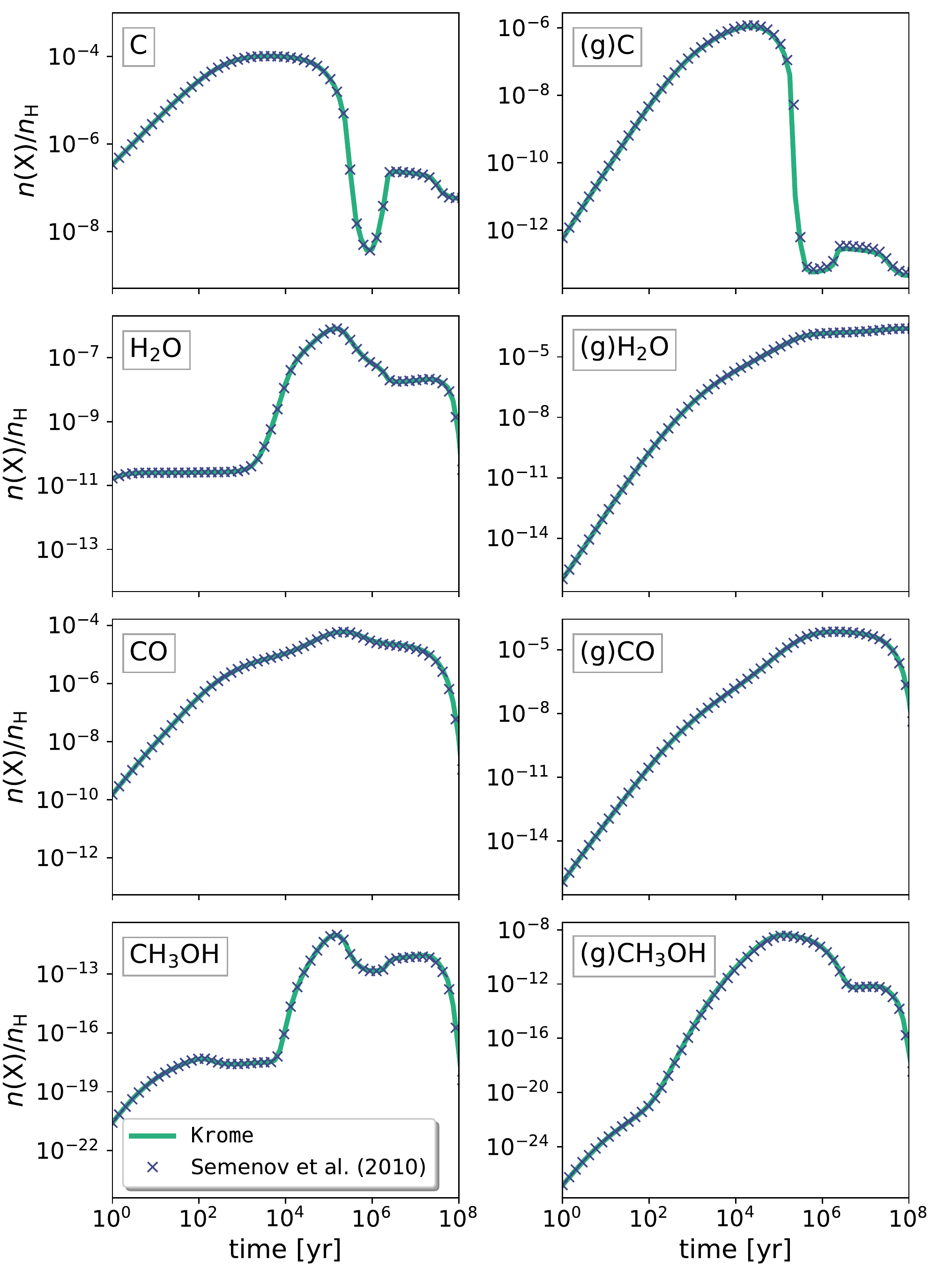}
\centering
\caption{Comparison of the time-dependent variation in the abundances of four arbitrary species (i.e. C, H$_2$O, CO, and CH$_3$OH) in gas phase ({\it left panels}) and on dust ({\it right panels}) in the TMC1 model. The green line shows the results obtained with the \krome -package, and the blue crosses are the results of \citet{Semenov10} with {\sc Alchemic}.}
\label{fig:benchmark1}
\end{figure*}

\begin{table}
\caption{{Summary of fiducial initial elemental abundances} set for the TMC1 model.}
\setlength{\tabcolsep}{10pt}
\renewcommand{\arraystretch}{1.3}
\centering
\begin{tabular}{lc|lc}
	\toprule
	\toprule
        species 	  	&  ($n_i/n_{\rm H}$)$_{\rm t=0}$ & species 	  	&  ($n_i/n_{\rm H}$)$_{\rm t=0}$ 	\\
    \midrule
 	He 		& 9.00(-2)	& Si$^+$ 	& 8.00(-9) \\
 	H$_2$ 	& 5.00(-1)	& Na$^+$ 	& 2.00(-9) \\
 	C$^+$ 	& 1.20(-4)	& Mg$^+$ 	& 7.00(-9)\\
 	N 		& 7.60(-5)	& Fe$^+$ 	& 3.00(-9)\\
 	O 		& 2.56(-4)	& P$^+$ 	& 2.00(-10)\\
 	S$^+$ 	& 8.00(-8)	& Cl$^+$ 	& 1.00(-9)\\
\bottomrule
\bottomrule
\end{tabular}\label{Tab:ICC_seme}
\tablefoot{{Same notation as in Tab.~\ref{Tab:ICC_PH1}.}}
\end{table}

We benchmarked our network against \cite{Semenov10}. The initial conditions are summarised in Tab.~\ref{Tab:ICC_seme}. All the elements are initially atomic, with the exception of hydrogen, which was assumed to be completely in molecular form. Except for He, N, and O, all the elements are also ionised, and grains are initially neutral.\\
\begin{figure*} 
        \includegraphics[width=1\textwidth]{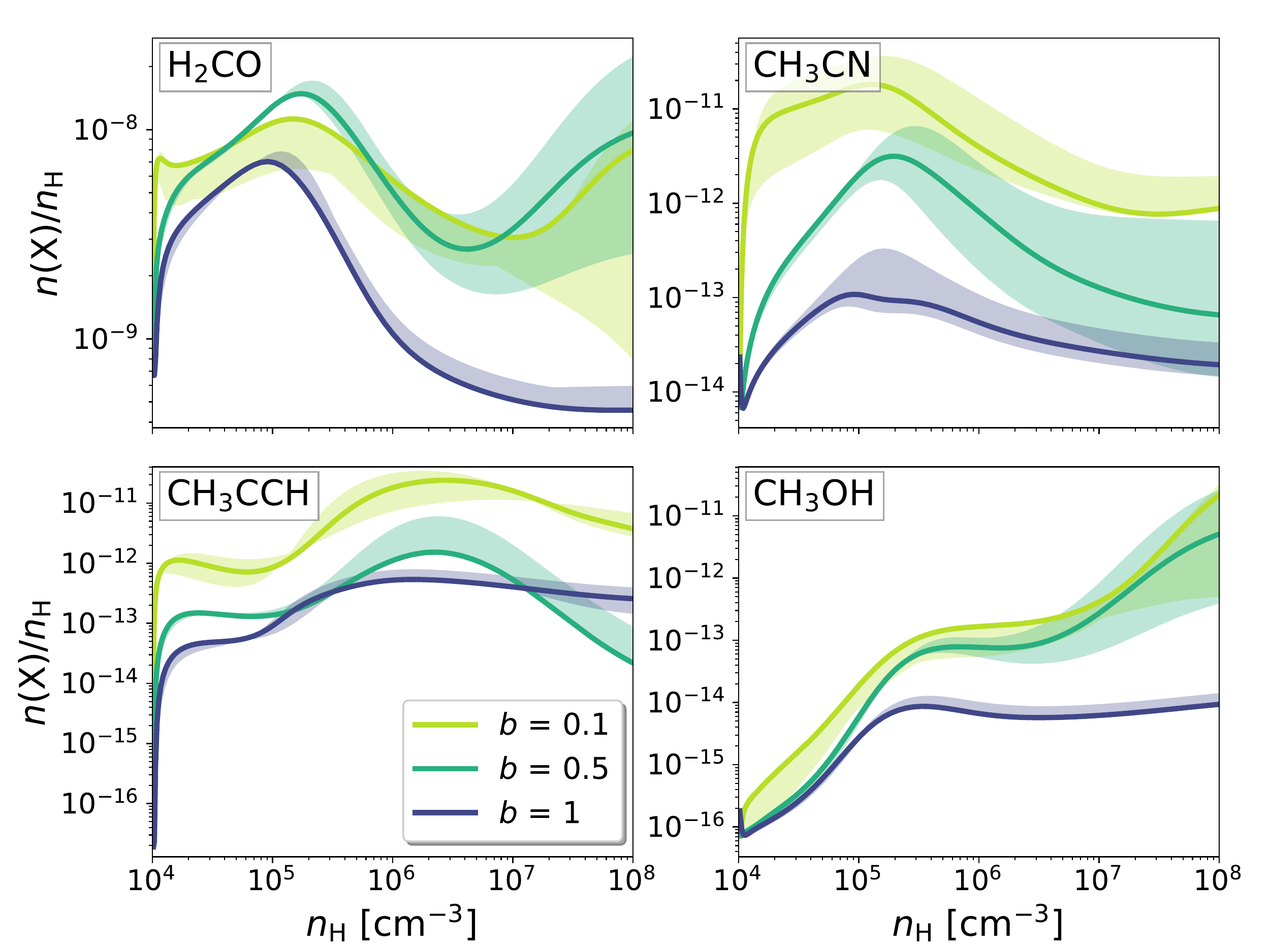}
        \centering
        \caption{Fractional abundance evolution for \hhco, \chhhcn, \chhhcch,~and \chhhoh~as a function of the central density of the clump during the collapse phase. Colours indicate the value of $b$ assumed for the collapse described by Eq.~\ref{eq:modi_tff}. Solid lines show the results employing the canonical binding energies from KIDA, and shaded areas represent the solutions when $T_b$ was varied by $\pm10\%$.}
        \label{fig:collapseprofs}
\end{figure*}

\noindent {\bf TMC1-model:} We benchmarked the physical case called TMC1 model. The gas has a constant temperature of 10 K, a constant hydrogen nuclei density $n_{\rm H}= 2\times10^{4}$ cm$^{-3}$, and visual extinction $A_{\rm v} = 10$ mag. The sticking coefficient was $S=1$, and the cosmic-ray ionisation rate of hydrogen molecules was set to $\crir = 1.3 \times 10^{-17}$ s$^{-1}$ \citep{Spitzer68, Glassgold74}. We have assumed an average grain size of $\left\langle a\right\rangle = 0.1~\mu$m and a bulk density $\rho_0=3$ g cm$^{-3}$ , corresponding to silicate grains, as in Sect.~\ref{sec32:methods_phys_model}. Figures~\ref{fig:benchmark1} shows the perfect agreement with the results reported by \cite{Semenov10}.

\section{Chemical evolution and model uncertainties during phase I}\label{app:PH1}

Fig.~\ref{fig:collapseprofs} shows the fractional abundance evolution of the relevant tracers (see Sect.~\ref{sec1:intro}) as a function of the central density of the clump during the isothermal collapse. We first tested  the effect of the collapse velocity (parameter $b$ in Eq.~\ref{eq:modi_tff}) on the chemical evolution of some of the observed species by delaying the collapse by 50\% and the 90\%, that is, $b=0.5$ and $b=0.1$, respectively. Different colours in Fig.~\ref{fig:collapseprofs} represent different values of $b$.\\
\indent Despite the effects induced by varying $b$, the relative importance of the chemical reactions involving some of the main tracers remains unaltered. In particular, \hhco~is mainly formed  through ${\rm CH_3} + {\rm O} \rightarrow {\rm H_2CO} + {\rm H}$, and it is destroyed by ion-molecule reactions that involve H$^+$, C$^+,$ and S$^+$. For densities higher than $\sim$10$^5$ cm$^{-3}$, ion-molecule reactions become predominant and the abundance of \hhco~decreases. At higher densities, some relative abundance profiles again increase because the dominant reaction becomes ${\rm H_3CO^+} + {\rm e^-} \rightarrow {\rm H_2CO} + {\rm H}$.\\
\indent Methyl acetylene is formed by the dissociative recombination  ($i$) ${\rm C_3H_5^+} + {\rm e^-} \rightarrow {\rm CH_3CCH} + {\rm H}$, the main gas-phase source of \chhhcch~(see \citealt{Hickson16}). For densities above $\sim$10$^6$ cm$^{-3}$, the relative abundance is almost constant because \chhhcch~reacts with C with a rate coefficient similar to ($i$).\\
\indent Conversely, the abundance of acetonitrile is more sensitive to $b$. The main source of gas-phase \chhhcn~in our model is ($ii$) ${\rm HCN} + {\rm CH_3^+} \rightarrow {\rm CH_3CNH^+}$, followed by the dissociative recombination ($iii$) ${\rm CH_3CNH^+} + e^-\rightarrow {\rm CH_3CN + H}$. We found that at beginning of the collapse, \chhhcn~ is rapidly formed, in agreement with what was estimated by \cite{Loison14}, and is then slowed down by ($iv$) ${\rm CH_3CN} + {\rm H_3^+} \rightarrow {\rm CH_3CNH^+} + {\rm H_2}$. The combined effect of reactions ($iii$) and ($iv$) is also sensitive to $b$ because for fast collapse ($b=1$), reactions ($ii$) and ($iii$) have less time to form \chhhcn~before reaction ($iv$) becomes effective. For densities between 10$^5$ and 10$^6$ cm$^{-3}$, \chhhcn~starts to react efficiently with ${\rm H}_3^+ \rightarrow {\rm CH_3CNH^+} + {\rm CO}$. The effect of this reaction in combination with ($iv$) produces a slow decrease in the \chhhcn~profiles at high densities.\\
\indent The abundance of methanol increases with density, driven by reaction ${\rm CH_3^+} + {\rm H_2O} \rightarrow {\rm CH_3OH_2^+}$ , followed by ${\rm CH_3OH_2^+ + e^- \rightarrow CH_3OH + H}$ (see also \citealt{Viti99}). 

To determine other sources of uncertainties, we analysed the variation in chemical evolution by changing the binding energies reported in Tab.~\ref{Tab:bindings} by $\pm10\%$, repeating the collapse with the same physical conditions. The effects on the fractional abundances are represented by the coloured areas in Fig.~\ref{fig:collapseprofs}. The discrepancy is more than one order of magnitude at the highest densities. We note that even larger discrepancies are reported by \cite{Penteado17}, when the binding energies are randomly selected from a normal distribution (see results reported in their Fig.~2).\\
\indent These tests show that the uncertainties due to {binding energies} variation are not negligible (see Fig.~\ref{fig:collapseprofs}), but they are not greater than variations caused by variations in $b$. This suggests that the dynamical state of the clump may play an important role in regulating its chemical evolution (see also the recent results by \citealt{Kulterer20}).\\
\indent Finally, we tested our collapse-phase with a set-up similar to that of \cite{Garrod06}, but employing the chemical network and binding energies of \cite{Semenov10} because both are publicly available and consequently easy to benchmark. The difference between the abundances found by \cite{Garrod06} and ours is within an order of magnitude. It is caused by the adopted details of the chemical networks.

\section{Notes on the observed column densities}\label{app:Nobs}

In Sect.~\ref{sec43:column-densities} we discussed that we post-processed the outputs of the models to compare the synthetic column densities with a number of available observations of the TOP100 clumps (right panel of Fig.~\ref{fig:sketch}). In this appendix we report some notes about our treatment of the observed column densities before the comparison that we discussed in Sect.~\ref{subsec41:comparison}. The observed column densities we used in the comparison are summarised in Tab.~\ref{Tab:observed-prop}.

\subsection{Formaldehyde {\rm (\hhco)}}
Formaldehyde data were taken from \cite{Tang18}. For simplicity, we assumed a single {half power beam width (HPBW)} size of 29\arcsec~to convolve the modelled column densities (i.e. the APEX-telescope resolution corresponding to the $J = 3-2$ transition of the \hhco~$ortho$ and the $para$ forms at $\sim$211 and $\sim$218 GHz, respectively). For the sources in which both isomers are observed, the total column density, $N_{\rm obs}$(\hhco), is derived as  $N$($o$-\hhco)$+N$($p$-\hhco). For sources not detected in $o$- and/or $p$-\hhco, we estimated the corresponding detection limit column density using \textsc{Weeds} (\citealt{Maret11}), with a 3$\sigma$ detection with respect to the average noise level (\citealt{Tang18}; 0.06 K main-beam temperature scale) and assuming the \hhco~molecular line parameters provided by the CDMS\footnote{\url{https://cdms.astro.uni-koeln.de/cdms/portal/}} database (\citealt{Muller01}). For each evolutionary class, we assumed the average values of $T_{\rm kin}$ (i.e. 52, 73, 81, and 110 K for 70w, IRw, IRb, and HII, respectively) and a {full width at half maximum (FWHM)} of 5 km s$^{-1}$ for the line reported by \cite{Tang18}, and that the source size was equal to the APEX beam at the corresponding frequency.

\subsection{Methyl acetylene {\rm (\chhhcch)}, methanol {\rm (\chhhoh),} and acetonitrile {\rm (\chhhcn)}}
Column densities of methyl acetylene, methanol, and acetonitrile were taken from \cite{Giannetti17_june}, who observed the $J = 20_{\rm K}-19_{\rm K}$, $J = 7_{\rm K}-6_{\rm K}$, and $J = 19_{\rm K}-18_{\rm K}$ bands for \chhhcch, \chhhoh, and \chhhcn, respectively, with APEX. {Because} these lines are very close to each other, we assumed the same angular resolution for APEX, that is, 18\arcsec. The final column densities in \cite{Giannetti17_june} were corrected for the APEX-beam dilution, $\eta = \theta_{\rm S}^2/(\theta_{\rm S}^2+\theta_{\rm beam}^2)$, where $\theta_{\rm S}$ and $\theta_{\rm beam}$ are the source and beam {HPBW} sizes, {respectively}. We (re-)applied this correction to each source where the detections were not resolved in the APEX-beam angular size (i.e. when $\theta_{\rm S}<\theta_{\rm beam}$) to obtain the beam-averaged column densities, whereas for sources with $\theta_{\rm S}>\theta_{\rm beam}$, $\eta=1$. Detection limits were estimated assuming the mean noise of the source spectra, the average line width, and the median excitation temperatures reported in Tables 6 and 7 of \cite{Giannetti17_june} for each evolutionary class. We treated the observed hot components for these tracers (see Sect.~\ref{sec43:column-densities}) in the same way as the colder ones, considering the appropriate temperatures reported by \cite{Giannetti17_june}, when we calculated the detection limits.

\begin{table*}
	\caption{Summary of the observed properties of the selected TOP100 sources.}
	\setlength{\tabcolsep}{1.5pt}
    \renewcommand{\arraystretch}{1.2}
	\small
	\centering
	\begin{tabular}{lccccccccccc}
		\toprule
		\toprule
$\:\:\:\:\:\:\:$(1) & (2) & (3) & (4) & (5) & (6) & (7) & (8) & (9) & (10) & (11) & (12) \\
Source-ID    &Class & d$_{\odot}$ & $L_{\rm bol}$ & $M_{\rm clump}$& $T_{\rm d}$ & 
\multicolumn{6}{c}{log$_{10}$[$N_{\rm obs}$(\X/cm$^{-2}$)]}\\
\cmidrule(rl){7-12}
 & & kpc & $10^2\:L_\odot$ & $10^2\:M_\odot$ &   K  &\hhco &\chhhcch &(\chhhcn)$_{\rm cold}$ &(\chhhcn)$_{\rm hot}$ & (\chhhoh)$_{\rm cold}$ & (\chhhoh)$_{\rm hot}$ \\
\midrule
G13.18+0.06  & 70w &  2.40 &    83.18 &   3.68 & 24.24 &\:\:\:13.66&\:\:\:14.62 &\:\:\:13.41 & $<$12.89 &\:\:\:16.02 & $<$14.22\\ 
G14.49-0.14  & 70w &  3.87 &     7.50 &  19.16 & 12.39 & $<$12.68  &\:\:\:14.52 &\:\:\:13.22 & $<$12.89 &\:\:\:15.85 & $<$14.22\\ 
G30.89+0.14  & 70w &  4.90 &     4.96 &  19.18 & 11.44 & $<$12.68  &\:\:\:14.16 &\:\:\:13.08 & $<$12.93 &\:\:\:15.29 & $<$14.17\\ 
G351.57+0.77 & 70w &  1.34 &     4.33 &   1.64 & 17.04 & $<$12.68  &\:\:\:14.15 &  $<$14.53 & $<$12.72 & $<$15.42 & $<$14.20\\ 
G353.42-0.08 & 70w &  6.06 &    45.08 &  17.91 & 17.12 & $<$12.68  & $<$15.73 &\:\:\:12.74 & $<$12.74 & $<$15.42 & $<$14.20\\ 
G354.95-0.54 & 70w &  1.91 &     4.84 &   1.47 & 19.13 & $<$12.68  & $<$15.73 &\:\:\:12.71 & $<$12.70 & $<$15.42 & $<$14.20\\ 
\midrule
G08.68-0.37  & IRw &  4.78 &   275.00 &  14.81 & 24.18 &\:\:\:13.72&\:\:\:15.11 &\:\:\:13.58 & $<$12.85 &\:\:\:15.22 & $<$14.41\\ 
G08.71-0.41  & IRw &  4.78 &     5.04 &  16.62 & 11.81 & $<$12.80  &\:\:\:14.46 &\:\:\:12.99 & $<$12.84 &\:\:\:15.19 & $<$14.36\\ 
G10.45-0.02  & IRw &  8.55 &   113.14 &  16.06 & 20.70 & $<$12.80  &\:\:\:14.36 &\:\:\:13.34 & $<$12.91 &\:\:\:14.96 & $<$14.53\\ 
G14.11-0.57  & IRw &  2.57 &    31.77 &   3.53 & 22.35 &\:\:\:13.46&\:\:\:14.63 &\:\:\:13.14 & $<$12.87 &\:\:\:15.37 & $<$14.40\\ 
G14.63-0.58  & IRw &  1.83 &    27.78 &   2.54 & 22.54 &\:\:\:13.54&\:\:\:14.67 &\:\:\:13.22 & $<$12.83 &\:\:\:15.82 & $<$14.41\\ 
G24.63+0.17  & IRw &  7.72 &    50.36 &  15.28 & 18.12 &\:\:\:13.20&\:\:\:14.08 &\:\:\:13.21 & $<$12.89 &\:\:\:15.03 & $<$14.35\\ 
G28.56-0.24  & IRw &  5.45 &    17.67 &  54.14 & 11.73 & $<$12.80  &\:\:\:14.58 &\:\:\:12.96 & $<$12.84 &\:\:\:14.97 & $<$14.32\\ 
G305.19-0.01 & IRw &  3.80 &   125.29 &   5.18 & 26.07 &\:\:\:13.52& $<$15.30 &\:\:\:13.33 & $<$12.85 & $<$14.86 & $<$14.40\\ 
G317.87-0.15 & IRw &  2.95 &    16.41 &   3.62 & 19.26 &\:\:\:13.76&\:\:\:14.82 &\:\:\:13.51 & $<$12.84 &\:\:\:15.88 & $<$14.40\\ 
G318.78-0.14 & IRw &  2.78 &    64.00 &   3.59 & 24.92 &\:\:\:13.26&\:\:\:14.22 &\:\:\:13.28 & $<$12.92 &\:\:\:15.26 & $<$14.39\\ 
G326.99-0.03 & IRw &  3.95 &    11.44 &   4.44 & 17.95 &\:\:\:13.84&\:\:\:14.88 &\:\:\:13.57 & $<$12.89 &\:\:\:15.65 & $<$14.39\\ 
G331.71+0.60 & IRw & 10.53 &   373.43 &  51.12 & 21.00 &\:\:\:13.86&\:\:\:14.74 &\:\:\:13.48 & $<$12.93 &\:\:\:16.00 & $<$14.42\\ 
G336.96-0.23 & IRw & 10.91 &    36.25 &  24.22 & 15.77 &\:\:\:13.36&\:\:\:14.71 &\:\:\:13.21 & $<$12.89 &\:\:\:15.54 & $<$14.40\\ 
G337.26-0.10 & IRw & 11.00 &   300.11 &  31.87 & 21.73 &\:\:\:13.35&\:\:\:14.40 &\:\:\:13.27 & $<$12.93 &\:\:\:15.34 & $<$14.43\\ 
G342.48+0.18 & IRw & 12.55 &   641.50 &  49.13 & 23.61 &\:\:\:13.47&\:\:\:14.44 &\:\:\:13.18 & $<$12.81 &\:\:\:15.15 & $<$14.39\\ 
G353.07+0.45 & IRw &  0.86 &     0.57 &   0.18 & 17.79 & $<$12.80  &\:\:\:14.55 &  $<$13.48 & $<$12.70 &\:\:\:15.11 & $<$14.46\\ 
\midrule
G34.41+0.23  & IRb &  1.56 &    48.39 &   2.14 & 26.13 &\:\:\:14.01&\:\:\:15.20 &\:\:\:14.78 &\:\:\:14.72 &\:\:\:16.15 &\:\:\:15.92\\ 
G35.20-0.74  & IRb &  2.19 &   235.80 &   4.63 & 29.54 &\:\:\:14.03&\:\:\:14.94 &\:\:\:14.26 & $<$14.14   &\:\:\:16.33 &\:\:\:15.19\\ 
G59.78+0.07  & IRb &  2.16 &    97.71 &   2.55 & 28.19 & $<$12.88  &\:\:\:14.36 &\:\:\:13.26 & $<$13.11   &\:\:\:15.21 & $<$14.59\\ 
G305.56+0.01 & IRb &  3.80 &   517.18 &   4.06 & 33.41 & $<$13.72  & $<$15.00 &\:\:\:13.02 & $<$12.96     &\:\:\:15.37 & $<$14.55\\ 
G310.01+0.39 & IRb &  3.61 &   496.33 &   4.15 & 32.20 &\:\:\:13.46&\:\:\:14.43 &\:\:\:13.23 & $<$12.94   &\:\:\:14.91 & $<$14.53\\ 
G313.58+0.32 & IRb &  3.78 &    94.14 &   1.83 & 29.17 & $<$12.88  & $<$15.00 &  $<$13.30 & $<$12.78      &\:\:\:14.88 & $<$14.88\\ 
G316.64-0.09 & IRb &  1.19 &     9.94 &   0.18 & 30.63 & $<$12.88  &\:\:\:14.67 &\:\:\:13.60 &\:\:\:13.57 &\:\:\:15.20 &\:\:\:15.12\\ 
G333.31+0.11 & IRb &  3.60 &   107.33 &   4.27 & 25.85 &\:\:\:13.73&\:\:\:14.73 &\:\:\:13.52 & $<$13.01 &\:\:\:15.63 & $<$14.58\\ 
G339.62-0.12 & IRb &  3.01 &   149.96 &   3.20 & 28.66 &\:\:\:13.65&\:\:\:14.69 &\:\:\:13.51 & $<$12.94 &\:\:\:14.94 & $<$14.57\\ 
G340.75-1.00 & IRb &  2.76 &    76.82 &   2.13 & 27.07 &\:\:\:13.64&\:\:\:14.35 &\:\:\:13.19 & $<$12.97 &\:\:\:15.75 & $<$14.49\\ 
G341.22-0.21 & IRb &  3.67 &   162.23 &   4.90 & 27.03 &\:\:\:13.74&\:\:\:14.33 &\:\:\:13.62 & $<$12.91 &\:\:\:15.56 & $<$14.59\\ 
G345.51+0.35 & IRb &  2.25 &   432.67 &   4.24 & 32.70 &\:\:\:14.04&\:\:\:14.85 &\:\:\:13.66 & $<$12.94 &\:\:\:16.10 &\:\:\:16.05\\ 
G345.72+0.82 & IRb &  1.56 &    18.73 &   2.01 & 22.05 &\:\:\:13.28&\:\:\:14.79 &\:\:\:13.09 & $<$12.89 &\:\:\:15.21 & $<$14.53\\ 
G351.77-0.54 & IRb &  1.00 &   164.69 &   2.64 & 31.79 &\:\:\:14.82&\:\:\:15.92 &\:\:\:15.68 &\:\:\:15.63 &\:\:\:16.73 &\:\:\:16.68\\ 
G353.41-0.36 & IRb &  3.44 &  1272.01 &  34.83 & 28.26 &\:\:\:13.98&\:\:\:15.03 &\:\:\:13.59 & $<$12.91 &\:\:\:15.88 & $<$14.68\\ 
\midrule
G10.62-0.38  & HII &  4.95 &  4227.57 &  37.87 & 34.45 &\:\:\:14.68&\:\:\:15.43 &\:\:\:13.98 &\:\:\:13.34 &\:\:\:15.95 &\:\:\:15.21\\ 
G12.81-0.20  & HII &  2.40 &  2468.69 &  18.80 & 35.15 &\:\:\:14.19&\:\:\:15.36 &\:\:\:13.64 & $<$13.23 &\:\:\:15.32 & $<$14.54\\ 
G31.41+0.31  & HII &  4.90 &   689.42 &  30.63 & 26.32 &\:\:\:14.14&\:\:\:15.34 &\:\:\:15.45 &\:\:\:15.43 &\:\:\:16.41 &\:\:\:16.38\\ 
G34.40+0.23  & HII &  1.56 &    29.91 &   2.76 & 22.82 &\:\:\:13.93&\:\:\:14.85 &\:\:\:13.75 & $<$13.09 &\:\:\:16.04 & $<$14.54\\ 
G301.14-0.23 & HII &  4.40 &  2137.40 &  19.41 & 34.56 & $<$13.02  &\:\:\:15.13 &\:\:\:14.01 &\:\:\:13.76 &\:\:\:15.91 &\:\:\:15.78\\ 
G330.88-0.37 & HII &  4.16 &  1545.88 &  15.87 & 33.44 &\:\:\:14.26&\:\:\:15.19 &\:\:\:14.61 &\:\:\:14.49 &\:\:\:16.11 &\:\:\:16.01\\ 
G330.95-0.18 & HII &  9.32 & 13064.99 & 173.06 & 33.04 &\:\:\:14.45&\:\:\:15.42 &\:\:\:14.58 &\:\:\:14.44 &\:\:\:16.21 &\:\:\:16.07\\ 
G332.83-0.55 & HII &  3.60 &  2419.56 &  19.36 & 35.67 &\:\:\:14.25&\:\:\:15.30 &\:\:\:14.22 &\:\:\:14.08 &\:\:\:16.03 &\:\:\:15.90\\ 
G333.28-0.39 & HII &  3.60 &  1291.69 &  20.77 & 30.37 &\:\:\:13.80&\:\:\:14.95 &\:\:\:13.47 & $<$13.28 &\:\:\:15.76 & $<$14.56\\ 
G333.60-0.21 & HII &  3.60 & 12416.71 &  34.61 & 41.12 &\:\:\:14.23&\:\:\:15.47 &\:\:\:13.75 & $<$13.53 &\:\:\:14.80 & $<$14.59\\ 
G351.42+0.65 & HII &  1.34 &   398.57 &   4.63 & 33.38 & $<$13.02  &\:\:\:15.27 &\:\:\:14.85 & $<$14.72 &\:\:\:16.88 &\:\:\:16.85\\ 

\bottomrule
\bottomrule
	\end{tabular}\label{Tab:observed-prop}
	\tablefoot{{Column 1: TOP100 sources considered in this work; Columns from (2) to (6): physical properties of each source \citep{Konig17}; Columns from (7) to (12): final beam- and LOS-average column densities, obtained by applying the procedure described in Appendix~\ref{app:Nobs} to the values reported in \cite{Giannetti17_june} and \cite{Tang18}. Upper limits are marked with '$<$'.}}
\end{table*}

\end{appendix}
\end{document}